\documentclass[12pt]{article}
\pdfoutput=1
\usepackage{amssymb}
\usepackage{amsmath}
\usepackage{latexsym}
\usepackage[usenames]{color}
\usepackage{fancybox}
\usepackage{simplewick}
\usepackage{comment}
\usepackage{cite}
\usepackage{framed}
\usepackage{booktabs}                       
\usepackage{multirow,bigdelim}              
\usepackage{longtable}                      
\definecolor{shadecolor}{rgb}{0.9,0.9,0.95}
\usepackage{setspace}
\usepackage{graphicx} 
\usepackage[usenames,dvipsnames]{xcolor}
\usepackage[setpagesize=false,pagebackref=false, linktocpage, bookmarksopen=true, colorlinks=true, linkcolor=RoyalBlue,citecolor=Maroon,urlcolor=Maroon]{hyperref}

\def\beq{\begin{equation}}
\def\eeq{\end{equation}}
\def\pmatrix#1#2{\left( 
\begin{array}{#1}
#2\end{array} 
\right)}
\def\del {\partial}

\def\comma{\,,}
\def\period{\,.}

\def\Xint#1{\mathchoice
   {\XXint\displaystyle\textstyle{#1}}%
   {\XXint\textstyle\scriptstyle{#1}}%
   {\XXint\scriptstyle\scriptscriptstyle{#1}}%
   {\XXint\scriptscriptstyle\scriptscriptstyle{#1}}%
   \!\int}
\def\XXint#1#2#3{{\setbox0=\hbox{$#1{#2#3}{\int}$}
     \vcenter{\hbox{$#2#3$}}\kern-.5\wd0}}

\def\dashint{\Xint-}

\begin{document}
\thispagestyle{empty}

\renewcommand{\thefootnote}{\fnsymbol{footnote}}
\setcounter{page}{1}
\setcounter{footnote}{0}
\setcounter{figure}{0}
\begin{flushright}
\end{flushright}
\vspace{1cm}
\begin{center}
{\large \bf
Functional Equations and Separation of Variables\\ for Exact $g$-Function
\par}

\vspace{2cm}

\textrm{Jo\~{a}o Caetano$^{a,b}$, Shota Komatsu$^{c}$}
\\ \vspace{1cm}
\footnotesize{\textit{
$^{a}$C. N. Yang Institute for Theoretical Physics, SUNY, Stony Brook, NY 11794-3840, USA\\
$^{b}$Simons Center for Geometry and Physics, SUNY, Stony Brook, NY 11794-3636, USA\\
$^{c}$School of Natural Sciences, Institute for Advanced Study, Princeton, NJ 08540, USA
}  
\vspace{1cm}
}

{\tt  joao.caetanus AT gmail.com, skomatsu AT ias.edu}

\par\vspace{2cm}

\textbf{Abstract}\vspace{2mm}
\end{center}
\noindent
The $g$-function is a measure of degrees of freedom associated to a boundary of two-dimensional quantum field theories. In integrable theories, it can be computed exactly in a form of the Fredholm determinant, but it is often hard to evaluate numerically. In this paper, we derive functional equations---or equivalently integral equations of the thermodynamic Bethe ansatz (TBA) type---which directly compute the $g$-function in the simplest integrable theory; the sinh-Gordon theory at the self-dual point. The derivation is based on the classic result by Tracy and Widom on the relation between Fredholm determinants and TBA, which was used also in the context of topological string.  We demonstrate the efficiency of our formulation through the numerical computation and compare the results in the UV limit with the Liouville CFT. As a side result, we present multiple integrals of $Q$-functions which we conjecture to describe a universal part of the $g$-function, and discuss its implication to integrable spin chains.

\setcounter{page}{1}
\renewcommand{\thefootnote}{\arabic{footnote}}
\setcounter{footnote}{0}
\setcounter{tocdepth}{2}
\newpage
\tableofcontents
\parskip 5pt plus 1pt   \jot = 1.5ex
\section{Introduction\label{sec:intro}}
\paragraph{General motivation} In two-dimensional quantum field theories, there exists a function called the $c$-function \cite{Zamolodchikov:1986gt} which provides a measure of degrees of freedom in the theory. Two important properties of the $c$-function are that it monotonically decreases along the RG flow and that it is stationary at fixed points and coincides with the conformal anomaly.

In \cite{affleck1991universal}, Affleck and Ludwig introduced an analogue of the $c$-function for theories in two dimensions with a boundary. The quantity, called the $g$-function or the boundary entropy, is defined as an overlap between the boundary state and the ground state. Although it may not be obvious from its definition, the $g$-function is a measure of degrees of freedom associated to a boundary. A justification of this claim comes from its monotonicity along the RG flow, which was proven in \cite{Friedan:2003yc}\footnote{A slightly different quantity, which is also monotonic along the RG flow, was introduced in \cite{Casini:2016fgb}. It coincides with the $g$-function at fixed points but is different along the RG flow.}.
 
 Given its importance, it is interesting to compute the $g$-function explicitly. However, except at critical points where one can use the techniques of conformal field theories (CFTs), the computation of the $g$-function is generally difficult. This motivates one to study a class of theories which are exactly solvable along the RG flow; namely integrable field theories.
   
 The computation of the $g$-function in integrable theories has a rather convoluted history: The first attempt to compute the $g$-function was made in \cite{LeClair:1995uf}. However, the paper \cite{Woynarovich:2004gc} pointed out the incompleteness of the result in \cite{LeClair:1995uf} and proposed a modified formula. Later a more systematic analysis was carried out  in \cite{Dorey:2004xk} and a further modification to the proposal was made.  This result was verified in \cite{Pozsgay:2010tv}, which re-derived it from a different approach. More recently, the paper \cite{Kostov:2018dmi} performed a rigorous analysis and confirmed the result in \cite{Dorey:2004xk} and extended it to theories with non-diagonal scatterings. This analysis was beautifully reformulated in \cite{Kostov:2019sgu} in terms of a path integral of an effective quantum field theory which localizes to the saddle point. This streamlines the derivations and clarifies the origin of the simplicity of the final answer. As another extension, the ``excited-state $g$-function''---namely the overlap between the boundary state and an excited state---was discussed in \cite{Kostov:2018dmi} and a complete formula was written down in \cite{Jiang:2019xdz,Jiang:2019zig}.
 
 The results obtained in these works are given in a form of the Fredholm determinant. Although they provide a closed-form expression amenable to numerical computations, the evaluation of the Fredholm determinant is often computationally costly. The main aim of this paper is to reformulate the $g$-functions in the simplest integrable theory (the sinh-Gordon theory) into a more efficient form---namely functional equations or equivalently integral equations of the Thermodynamic Bethe Ansatz (TBA) type. The derivation is based on the result by Tracy and Widom in \cite{Tracy:1995ax} and its generalization in \cite{Okuyama:2015pzt}, which showed that certain Fredholm determinants can be computed by solving the TBA-like equations. We demonstrate the efficiency of this reformulation through the numerical computation. As a side result, we write down multiple integrals which we conjecture to describe a universal part of the $g$-function in the sinh-Gordon theory. They are given in terms of the so-called $Q$-functions and resemble integrals obtained in the {\it separation of variables} approach \cite{Sklyanin:1995bm}. 
 \paragraph{$g$-Functions in integrable theories}
The $g$-function is important also for the study of integrability itself. This is because the $g$-function is one of the rare quantities beyond the spectrum which can be computed exactly at finite volume. 
 
 In integrable field theories, there is by now a textbook prescription for computing the spectrum: The first step is to consider a theory in the infinite volume and determine the exact S-matrix using the symmetry, the Yang-Baxter equation and other consistency conditions. In the second step, one writes down the asymptotic Bethe ansatz equations which are basically the quantization conditions for individual excitations. The asymptotic Bethe ansatz is known to capture all the perturbative $1/L$ corrections in a large-volume ($L$) expansion. To include the nonperturbative corrections (which are of the form $e^{-L \bullet}$), one needs to Wick-rotate and swap the roles of space and time. This maps the computation of the {\it finite-volume} spectrum to the computation of the free energy at {\it finite temperature}. One can then use the techniques called the thermodynamic Bethe ansatz and determine the exact finite-volume spectrum. 
 
 Unfortunately, the analogue of such a prescription is not known for most of other quantities. For instance, the form factor of local operators, which might seem like the next step in complexity, is not known\footnote{Although there are some results for two-particle form factors, the results for multiparticle form factors are not known in general.} in general even in the infinite volume. One of the difficulties comes from the fact that they are ``$O(1)$ quantities'': While the spectrum (in particular the ground-state energy) can be read off easily from the asymptotic behavior of the Wick-rotated partition function $Z\sim e^{-R E}$, the computation of other observables requires us to determine the $O(1)$ piece of the partition function or the partition function with operator insertions. (See section \ref{sec:review} for more details.)
 
One of the rare exceptions\footnote{Other important exceptions are the vacuum one-point functions and the finite-volume diagonal form factors in sine-Gordon and sinh-Gordon theories \cite{Jimbo:2009ja,Jimbo:2010jv,Negro:2013wga,Bajnok:2019yik,Hegedus:2019rju}. However, these results rely heavily on specific properties of these theories and do not seem as universal as $g$-functions.} is the $g$-function, for which one can generalize the analysis for the spectrum and perform the computation exactly and systematically. In this respect, the $g$-function stands at a unique position in the study of integrability; it is an $O(1)$ quantity much like other observables but nevertheless exactly computable in the finite volume. For this reason, it would be fruitful to further develop the techniques to compute the $g$-functions in integrable theories and see how far we can get.
 
 \paragraph{Relation to $\mathcal{N}=4$ supersymmetric Yang-Mills} Yet another motivation comes from planar $\mathcal{N}=4$ supersymmetric Yang-Mills theory ($\mathcal{N}=4$ SYM) in four dimensions. It was found more than a decade ago \cite{Minahan:2002ve} that the computation of the spectrum of local operators in $\mathcal{N}=4$ SYM can be mapped to the computation of the energy in integrable spin chains. Following the ``textbook'' prescription mentioned above, a complete determination of the spectrum was achieved\footnote{A more elegant and efficient formalism called the quantum spectral curve was developed in \cite{Gromov:2013pga,Gromov:2014caa}.} in \cite{Gromov:2009tv,Gromov:2009bc,Gromov:2009zb,Bombardelli:2009ns,Arutyunov:2009ur}. 
 
 A natural next step along this line is to compute correlation functions of local operators. It was found in \cite{Basso:2015zoa,Fleury:2016ykk,Eden:2016xvg,Bargheer:2017nne,Eden:2017ozn,Bargheer:2018jvq} that they can be decomposed into basic building blocks called the {\it hexagon form factors}, which were determined in \cite{Basso:2015zoa} exactly. However, to actually compute correlation functions of operators of finite lengths, one needs to sum over all possible intermediate states which appear in the hexagon-form-factor expansion. This leads to an infinite series which is in practice hard to evaluate. 
 
 More recently, it was found in \cite{Jiang:2019xdz,Jiang:2019zig} that a special class of three-point functions can be computed exactly using the techniques of the $g$-functions. The result is given in a closed form as the Fredholm determinant and is by far more compact and practical than the infinite series in the hexagon approach. Nevertheless, the Fredholm determinant is still rather complicated to evaluate and it would be worth seeking a further reformulation. Although the results in this paper cannot be applied directly to $\mathcal{N}=4$ SYM, we think that they provide the first step towards such a goal.

 \paragraph{Structure of the paper} The rest of this paper is organized as follows. In section \ref{sec:review}, we review the derivation of the Fredholm-determinant formula for the $g$-function in integrable theories. Besides explaining the derivation for general integrable models, we also write down explicit formulae for the sinh-Gordon theory. In section \ref{sec:derivation}, we explain how to rewrite the $g$-functions in the sinh-Gordon theory into functional equations and the TBA-type integral equations. The section also contains a concise review of the results by Tracy and Widom on the relation between the Fredholm determinant and TBA. In section \ref{sec:numerics}, we use the TBA-type equations that we derived and compute the $g$-function numerically. The main purpose is to show the efficiency of our formulation and to compare the result in the small radius limit with the result computed by CFT. In section \ref{sec:sov}, we present  multiple integrals which we conjecture to describe a universal part of the $g$-function in the sinh-Gordon theory. The resulting expression resembles integrals in the separation of variables approach and generalizes the result by Lukyanov \cite{Lukyanov:2000jp} on the finite-volume one-point functions. We also mention possible applications of our argument to the overlaps between boundary states and Bethe states in integrable spin chains. 
 In section \ref{sec:conclusion}, we summarize the results and discuss the future directions. Technical details are often relegated to the appendices.

\section{g-Functions in integrable theories\label{sec:review}}
\begin{figure}[t]
\centering
\includegraphics[clip,height=2.5cm]{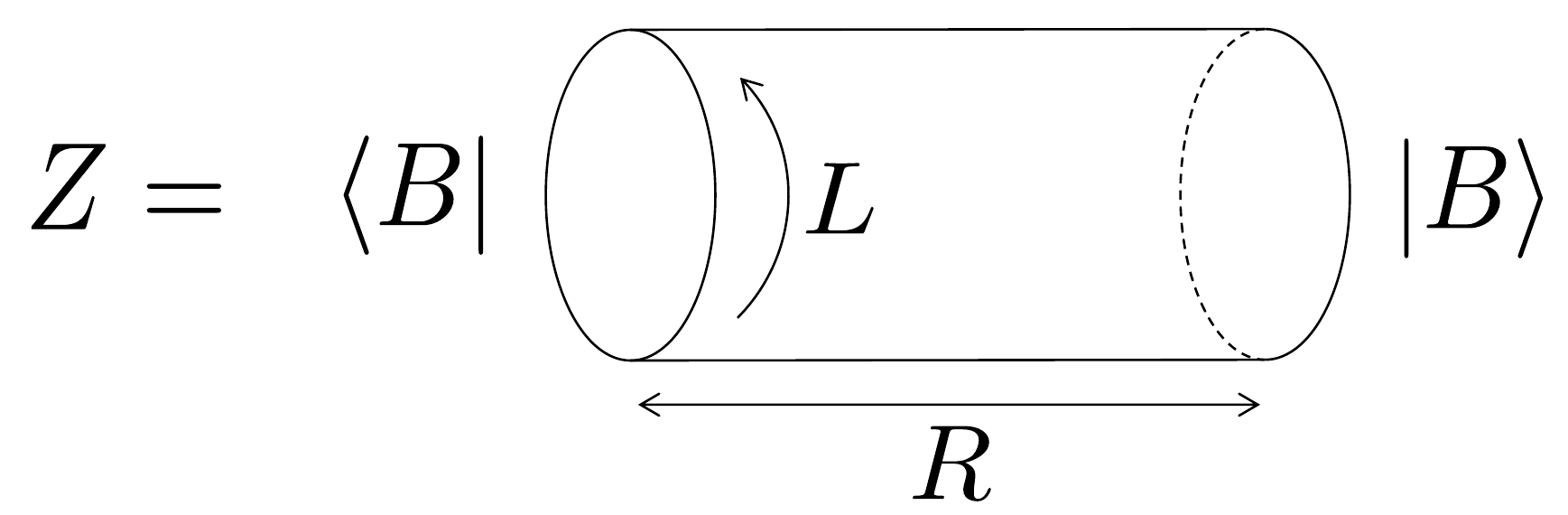}
\caption{The cylinder partition function with circumference $L$ and length $R$. It admits two different expansions; the closed string channel \eqref{eq:expandclosed} in which $R$ plays the role of time, and the open string channel \eqref{eq:expandopen} in which $L$ plays the role of time. Equating two expansions in the $R\to \infty$, we obtain the TBA formulation for the $g$-function \eqref{eq:finalgfunction}.} \label{fig:cylinder}
\end{figure}
\subsection{Review of the derivation of the $g$-function}
To discuss the $g$-function in integrable theories, we consider a partition function of a cylinder with circumference $L$ and length $R$ whose boundaries are contracted with boundary states\footnote{One can also choose the two boundary states to be different, $\langle B|$ and $|B^{\prime}\rangle$, but for simplicity here we take them to be the same.} $\langle B|$ and $|B\rangle$ (see figure \ref{fig:cylinder}). This partition function, which we denote by $Z$, can be computed in two different channels. The first one is the ``closed string'' channel in which we regard $R$ as the Euclidean time direction. Inserting a complete set of states $\psi_c$ on the circle of circumference $L$, we obtain an expansion of the partition function,
\beq\label{eq:expandclosed}
Z=\sum_{\psi_c}e^{-R E_{\psi_c} (L)} \frac{\langle B|\psi_c\rangle \langle \psi_c|B\rangle}{\langle \psi_c|\psi_c\rangle}\,,
\eeq
where $E_{\psi_c}(L)$ is the energy of the state $\psi$.
To isolate the contribution from the ground state $|\Omega\rangle$, we consider the limit $R\to \infty$. Then we can approximate the partition function as
\beq\label{eq:Zinclosed}
Z\sim e^{-R E_{\Omega}}|g|^2 \,, 
\eeq
where $g$ is the $g$-function defined by
\beq
g:=\frac{\langle B|\Omega \rangle}{\sqrt{\langle \Omega|\Omega\rangle}}\,.
\eeq

To compute the $g$-function using integrability, we need to consider the other channel; namely the ``open string'' channel in which we regard $L$ as the Euclidean time direction. Inserting a complete set of states $\psi_o$ on the segment of length $R$, the partition function can be expanded as
\beq\label{eq:expandopen}
Z=\sum_{\psi_o} e^{-LE_{\psi_o}(R)}\,,
\eeq
where $E_{\psi_o}$ is the energy of the state $\psi_o$. Equating the two expansions \eqref{eq:expandclosed} and \eqref{eq:expandopen} and taking the limit $R\to \infty$, we obtain
\beq\label{eq:gfunctionlimit}
e^{-R E_{\Omega}}|g|^2=\lim_{R\to\infty}\sum_{\psi_o}e^{-LE_{\psi_{o}}(R)}\,.
\eeq
Now the right hand side of \eqref{eq:gfunctionlimit} can be identified with the thermal partition function at temperature $1/L$ in the infinite volume limit ($R\to \infty$). To proceed, we furthermore assume that the boundary state $|B\rangle$ does not break integrability. In other words, we assume that $|B\rangle$ belongs to a class of {\it integrable boundary states} introduced in \cite{Ghoshal:1993tm}. In such a case, we can apply the standard arguments of the thermodynamic Bethe ansatz and compute the thermal partition function. Namely we replace the sum over states with a path integral of densities of particles $\rho$
\beq
\lim_{R\to\infty}\sum_{\psi_o}e^{-LE_{\psi_{o}}(R)} \sim \lim_{R\to \infty} \mathcal{N}\int D\rho \,  e^{-R S_{\rm eff}[\rho]}\,.
\eeq
As is evident from the expression, this path integral is dominated by the saddle point in the $R\to \infty$ limit. One should however note that the saddle-point value of the action $S[\rho^{\ast}]$ only gives the exponential piece in \eqref{eq:gfunctionlimit}, $e^{-R E_{\Omega}}$. In order to compute the $g$-function, one also needs to include the one-loop fluctuation around the saddle point and the $O(1)$ normalization constant $\mathcal{N}$. For details of the derivation, see for instance \cite{Pozsgay:2010tv,Kostov:2018dmi,Kostov:2019sgu} and section 6 of \cite{Jiang:2019xdz}.

Combining all the factors, one obtains a closed-form expression for the $g$-function. The result for a theory with a single species of particles without bound states reads \cite{Dorey:2004xk,Pozsgay:2010tv,Kostov:2018dmi,Jiang:2019xdz}\footnote{The result can be generalized to theories with multiple species of particles and with bound states as shown in \cite{Kostov:2018dmi,Jiang:2019xdz}. However, the resulting expression is divergent in general and one needs some regularization as pointed out in \cite{Kostov:2019fvw}. Nevetheless, the case relevant for the three-point function in $\mathcal{N}=4$ SYM seems free of divergences and the regularization seems unnecessary \cite{Jiang:2019xdz}.}
\beq\label{eq:finalgfunction}
\log g =\int^{\infty}_{0} 
\frac{du}{2\pi}\Theta (u) \log (1+e^{-\epsilon (u)}) +\frac{1}{2}\log \frac{{\rm Det}(1-\hat{G})}{\left({\rm Det}(1-\hat{G}_{+})\right)^2}\,.
\eeq
Here ${\rm Det}$'s are the Fredholm determinants and the operators $\hat{G}$'s are defined by
\beq
\hat{G}_{\pm}\cdot f(u):=\int^{\infty}_{0}\frac{dv}{2\pi} \frac{\mathcal{K}_{\pm}(u,v)}{1+e^{\epsilon (v)}}\,,\qquad \hat{G}\cdot f(u):=\int_{-\infty}^{\infty}\frac{dv}{4\pi}\frac{\mathcal{K}_{+}(u,v)+\mathcal{K}_{-}(u,v)}{1+e^{\epsilon (v)}} f(v)  \,,
\eeq
and the kernels $\mathcal{K}_{\pm}$ are given in terms of the S-matrix $S(u,v)$ as
\beq
\mathcal{K}_{\pm}(u,v)=\frac{1}{i}(\del_{u}\log S(u,v)\pm \del_{u}\log S(u,-v))\,.
\eeq
The only factor which is sensitive to the details of the boundary state is $\Theta (u)$, and is given in terms of the reflection factor $R(u)$ as
\beq
\Theta (u)=\frac{1}{i}\del_u \log R(u) -\pi \delta (u)-\frac{1}{i}\del_u \log S(u,v)|_{v=-u}\,.
\eeq
Finally, $\epsilon (u)$ is the pseudo-energy which satisfies the TBA equation
\beq
\epsilon (u)=L E(u)-\mathcal{K}_{+}\ast \log (1+e^{-\epsilon})\,,
\eeq
where $E(u) =m\cosh u$ is the energy of the excitation (in the open string channel) and $\ast$ is the convolution $A\ast B:=\int \frac{dv}{2\pi}A(u,v)B(v)$.

The result \eqref{eq:finalgfunction} in principle gives a nonperturbative expression for the $g$-function in integrable theories. However, the computation is rather complicated because of the Fredholm determinants. To evaluate the Fredholm determinants in practice, we either need to approximate them by finite-dimensional determinants or expand them in a series. For instance, the two determinants in the ratio can be combined into a single series given by
\beq \label{eq:seriesratio}
\log \frac{{\rm Det}(1-\hat{G})}{\left({\rm Det}(1-\hat{G}_{+})\right)^2} =  \sum_{n=1}^{\infty} \frac{1}{n}\int_{\mathbb{R}^n} \prod_{i=1}^{n} \frac{du_i}{2\pi} \frac{1}{1+e^{\epsilon(u_i)}} \mathcal{K}_s(u_1+u_2)\prod_{j=2}^{n} \mathcal{K}_s(u_j-u_{j+1}),
\eeq 
with $u_{n+1}\equiv u_1$ and $\mathcal{K}_s(u-v)=\frac{1}{i}\del_{u}\log S(u,v)$. 
The purpose of this paper is to derive an alternative expression for the $g$-function which is more suited for the numerical evaluation, and in particular, surpasses the multiple integrations.
\subsection{$g$-Functions in sinh-Gordon}
We will consider the simplest  interacting quantum field theory in two dimensions: the sinh-Gordon theory. In order to set up the notations, we will quickly review a few well-known facts about this model. The sinh-Gordon is an integrable theory defined by the Lagrangian
\beq \label{eq:SGlag}
\mathcal{L} = \frac{1}{4\pi} (\partial \phi)^2 + 2\mu \cosh (2b\phi)\,,
\eeq
containing two parameters: the coupling $b$ and the cosmological constant $\mu$. Its spectrum consists of a single massive particle whose mass $m$  is known in terms of the parameters in the Lagrangian (see \cite{Zamolodchikov:1995xk} and appendix \ref{app:liouville} for more details on this relation).
The corresponding scattering is encoded in the following S-matrix,
\beq \label{eq:Smatrix}
S(u,v) = \frac{ \sinh(u-v) - i \sin(\pi p)}{ \sinh(u-v) + i \sin(\pi p)}\,,
\eeq
where the parameter $p$ is related to the coupling constant $b$ by the relation $p =b^{2}(1+b^2)^{-1}$ and throughout this work we will focus on the self-dual point for which $b=1$.

We will consider this model defined on geometries with specific boundaries that preserve integrability. These integrable boundary conditions can be derived from the boundary data associated to the lowest breather of the sine-Gordon model by the appropriate analytic continuation. Namely, we have that the sine-Gordon reflection matrix was determined in \cite{Ghoshal:1993tm, Ghoshal:1993iq} and given by,
\beq \label{eq:refmatrix}
\mathcal{R}_{\text{sG}}(\theta|\eta,\vartheta,\lambda) = \frac{f\left(\theta, \frac{1}{2}\right) f\left(\theta, \frac{1}{2\lambda} +1\right) }{f\left(\theta, \frac{1}{2\lambda} +\frac{3}{2}\right)} \frac{f\left(\theta, \frac{\eta}{\pi \lambda}-\frac{1}{2}\right) f\left(\theta, \frac{i \vartheta}{\pi \lambda} -\frac{1}{2}\right) }{ f\left(\theta, \frac{\eta}{\pi \lambda}+\frac{1}{2}\right) f\left( \theta, \frac{i \vartheta}{\pi \lambda} +\frac{1}{2}\right) } \;\;\; f(\theta,x) \equiv \frac{\sinh\left( \frac{\theta}{2}+\frac{ i \pi x }{2}\right)}{\sinh\left( \frac{\theta}{2} - \frac{ i \pi x }{2}\right)}\comma
\eeq
from where we can obtain the sinh-Gordon \cite{Dorey:1997yg, Bajnok:2007ep} one by setting 
\beq
\mathcal{R}_{\text{ShG}}(\theta| \eta,\vartheta, p)=\mathcal{R}_{\text{sG}}(\theta| i\eta,\vartheta,-1/p)\,,
\eeq
where $\eta, \vartheta$ are parameters characterizing the boundaries. With these results, the $g$-function is fully determined through (\ref{eq:finalgfunction}).

\section{Tracy-Widom TBA \label{sec:derivation}}
Fredholm determinants often provide a representation for several physical quantities that go  well beyond the $g$-functions. Examples include two-point functions in some 2d integrable models \cite{Bernard:1994re}, the $S^3$ partition functions of  supersymmetric gauge theories \cite{Calvo:2012du,Hatsuda:2013oxa,Hatsuda:2012dt,Putrov:2012zi,Hatsuda:2012hm}, nonperturbative formulation of topological string \cite{Grassi:2014zfa},   certain correlation functions and amplitudes in $\mathcal{N}=4$ SYM \cite{Jiang:2016ulr,Bettelheim:2014gma,Basso:2017khq,Belitsky:2019fan,Kostov:2019stn,Kostov:2019auq,Belitsky:2020qrm,Basso:2020xts}, $\,\mathcal{N}=2$ supersymmetric index in two dimensions \cite{Cecotti:1992qh}, and the partition function of 2d polymers \cite{Fendley:1992jy,Zamolodchikov:1994uw}. 
In the last two cases, it was pointed out by Zamolodchikov \cite{Zamolodchikov:1994uw} that such determinants are unexpectedly related to the solution of a set of integral equations reminiscent of the Thermodynamic Bethe Ansatz. This connection 
was later proven by Tracy and Widom in \cite{Tracy:1995ax}. For these reasons, we will call this set of equations Tracy-Widom TBA\footnote{Since it was first conjectured by Al.~Zamolodchikov, it might be more appropriate to call it ``Zamolodchikov TBA''. This however is not helpful for the purpose of distinguishing it from the standard TBA, which was also derived by Al.~Zamolodchikov. We therefore decided to call it Tracy-Widom TBA in this paper.}. Once proven, the relation between Fredholm determinants and functional equations was largely explored and further improved with great success in the context of the evaluation of the topological string partition function. Here, we will build on the Tracy-Widom proof and generalize it to our current case.

\subsection{Functional Equations} 
In our context, we deal with Fredholm determinants of the operators $\hat{G}$ and $\hat{G}_{\pm}$, see (\ref{eq:finalgfunction}). Let us start by representing each of them as a formal power series in the kernel. For example, the determinant involving $\hat{G}$  can be written as 
\beq
\begin{aligned} \label{eq:detexpand}
{\rm Det}(1- z\,{\hat G}) &=\exp\left(- \sum_{n=1}^{\infty} \frac{z^n}{n} \int_{\mathbb{R}^n} \prod_{i=1}^{n} \frac{du_{i}}{2\pi} \,\frac{ \mathcal{K}_s(u_i,u_{i+1}) }{1+e^{\epsilon(u_i)}} \right) \\ 
& = \exp\left( -\sum_{n=1}^{\infty} \frac{z^n}{n} \int_{\mathbb{R}^n} \prod_{i=1}^{n} \frac{du_{i}}{2\pi} K_s(u_i,u_{i+1})  \right) \\
&\equiv \exp \left(-\sum_{n=1}^{\infty}\frac{z^n}{n} {\rm{tr}}\, K_{s}^{\ast n} \right)
\end{aligned}
\eeq
with $u_{n+1}=u_1$ and we have introduced the bookkeeping parameter $z$ that might be set to one to recover the original problem. We have also defined the \textit{dressed} kernel by
\beq 
K_s (u,v) \equiv \frac{ \mathcal{K}_s(u,v)}{ \sqrt{1+e^{\epsilon(u)}}\sqrt{1+e^{\epsilon(v)}}}\comma
\eeq 
and the $j$-th convolution of a kernel $K^{\ast j}$ as
\beq \label{eq:jconv}
K^{\ast j} = \underbrace{K \ast K \ast\cdots \ast K}_{j}\comma
\eeq
with $f\ast g \equiv \int dv f(u,v) g(v,w)$. 
Analogously, we can rewrite the Fredholm determinant involving $\hat{G}_{+}$ in terms of a dressed kernel defined in a similar way from $\mathcal{K}_{+}$, that we denote from now on by $K_{+}$.
The important property of these kernels that will make our analysis possible is that both can be brought to the following form after using the explicit expression of the $S$-matrix in (\ref{eq:Smatrix}),
\beq \label{eq:defrhoM}
K_{\star}(u,v) =  \frac{E_{\star}(u)\, E_{\star}(v) }{ M_{\star}(u) + M_{\star}(v)}\;\,, \text{ with }\star=+,s \comma
\eeq
where  we define for $\star=s$,
\begin{equation}
E_s(u) \equiv  \frac{\sqrt{2} \,e^{u}}{\sqrt{ 1+e^{\epsilon(u)}}}\quad   M_s(u) \equiv e^{2u}\comma
\end{equation}
and for the kernel $\star=+$,
\begin{equation}\label{eq:kernelplusEandM}
E_{+}(u)= \frac{E_s(u)+E_s(-u)}{2}\quad M_{+} (u) = \frac{M_s(u)+M_s(-u)}{2} \period
\end{equation}
The kernels of  type (\ref{eq:defrhoM}) are particularly interesting because they admit a recursive representation \cite{Tracy:1995ax} from which one can derive functional equations for them. We highlight in the main text some important steps of this derivation and leave the details to the appendix \ref{app:technical}. 

\paragraph{Lemma by Tracy and Widom}The starting step is the following lemma proven by Tracy and Widom \cite{Tracy:1995ax}, which decomposes the kernels (\ref{eq:defrhoM}) in terms of some auxiliary functions of a single argument $\phi_{\star \, j}(u)$,
\beq \label{eq:decomp}
K_{\star}^{\ast n}(u,v)=\frac{E_{\star}(u)E_{\star}(v)}{M_{\star}(u)+(-1)^{n-1}M_{\star}(v)}\sum_{l=0}^{n-1}(-1)^{l}\phi_{\star\,l} (u) \phi_{\star\,n-1-l}(v)\period
\eeq
Here $\phi_{\star\,j }$ are defined recursively by
\beq \label{eq:recphi}
\phi_{\star \, j}(u) = \frac{1}{E_{\star}(u)} \int_{-\infty}^{\infty} dv\, K_{\star}(u,v)\, E_{\star}(v) \phi_{\star \,j-1} (v)\quad\quad \phi_{\star\, 0}(u)=1,\quad \text{with}\; \star=+,s\comma
\eeq
or equivalently by
\beq \label{eq:recursion}
\phi_{\star j} (u) = \frac{1}{E_{\star}(u)} \int_{-\infty}^{\infty} dv\, K_{\star}^{\ast j} (u,v) E_{\star}(v)\comma
\eeq
where $K^{\ast j}$ is the $j$-th convolution defined in (\ref{eq:jconv}).
In order to show (\ref{eq:decomp}), we  introduce the bra-ket notations of quantum mechanics. We can then reexpress (\ref{eq:recursion}) as
\beq\label{eq:opdefphi}
|E_{\star}\phi_{\star j}\rangle =\hat{K}_{\star}^{j}|E_{\star} \rangle \comma
\eeq
where $|E_{\star} \phi_{\star j}\rangle$ and $|E_{\star} \rangle$ are kets which correspond to the following ``wave functions'',
\beq
\langle u |E_{\star} \phi_{\star j}\rangle =E_{\star}(u)\phi_{\star j} (u) \,,\qquad \langle u |E_{\star}\rangle =E_{\star}(u) \period
\eeq
Similarly, the ``wave functions'' for the bra states $\langle E_{\star}|$ are given by\footnote{Since $E(u)$ is real in our case, this coincides with the standard definition of the bra state in quantum mechanics. When $E(u)$ is complex, in the standard definition, we have $\langle E|u\rangle =\bar{E} (u)$ with $\bar{E}$ being the complex conjugate of $E$. Therefore in that case it is more natural to replace the right hand side of \eqref{eq:togetprojector} with $|E\rangle \langle \bar{E}|$.} 
\beq
\langle E|u\rangle =E(u)\period
\eeq

\paragraph{Proof of the lemma}Using this representation and the equation \eqref{eq:defrhoM}, one can show the following operator equation (here we drop the subindices to make the notation less cluttered, since the Lemma is analogous for both kernels):
\beq\label{eq:togetprojector}
\hat{M}\hat{K}+\hat{K}\hat{M}=|E\rangle\langle E|\comma
\eeq
 \beq
\hat{M}|u \rangle =M(u)|u\rangle\period
\eeq
Using \eqref{eq:togetprojector}, we can explicitly compute (anti)commutators of $\hat{M}$ and $\hat{K}^{n}$ for small $n$:
\beq
\begin{aligned}
\hat{M}\hat{K}^2-\hat{K}^2\hat{M}&=(\hat{M}\hat{K}+\hat{K}\hat{M})\hat{K}-\hat{K}(\hat{M}\hat{K}+\hat{K}\hat{M})\\
&=|E\rangle\langle E|\hat{K}-\hat{K}|E\rangle\langle E|\comma\\
\hat{M}\hat{K}^3+\hat{K}^3\hat{M}&=(\hat{M}\hat{K}^2-\hat{K}^2\hat{M})\hat{K}+\hat{K}^2(\hat{M}\hat{K}+\hat{K}\hat{M})\\
&=|E\rangle\langle E|\hat{K}^2-\hat{K}|E\rangle\langle E|\hat{K}+\hat{K}^2|E\rangle\langle E|\period
\end{aligned}
\eeq
From this, it should be clear what the pattern is. Namely the result for general $n$ is
\beq\label{eq:lemmatracy}
\hat{M}\hat{K}^{n}-(-1)^{n} \hat{K}^{n}\hat{M}=\sum_{l=0}^{n-1}(-1)^{l}\hat{K}^{l}|E\rangle\langle E| \hat{K}^{n-1-l}\period
\eeq
To prove this, we use the mathematical induction. We will not fully explain the proof since it is straightforward. The main step of the proof is the identity
\beq
\hat{M}\hat{K}^{n}-(-1)^{n} \hat{K}^{n}\hat{M}=\Big[\hat{M}\hat{K}^{n-1}-(-1)^{n-1} \hat{K}^{n-1}\hat{M}\Big]\hat{K}-(-1)^{n}\hat{K}^{n-1}\Big[\hat{M}\hat{K}+\hat{K}\hat{M}\Big]\period
\eeq

We can bring \eqref{eq:lemmatracy} into a more convenient form by sandwiching both sides with $\langle u |$ and $|v\rangle$. We then get
\beq
\begin{aligned}
(\text{l.h.s})&=\left(M(u)+(-1)^{n-1}M(v)\right)K^{\ast n}(u,v)\comma\\
(\text{r.h.s})&=\sum_{l=0}^{n-1}(-1)^{l}\langle u |\hat{K}^{l}|E\rangle \langle E|\hat{K}^{n-1-l}|v\rangle \period
\end{aligned}
\eeq
Using the reality of $E$ and the kernel $K$ and the definition of $\phi_{\star j}$ given in \eqref{eq:opdefphi}, the right hand side can be re-expressed as
\beq
(\text{r.h.s})=E(u)E(v)\sum_{l=0}^{n-1}(-1)^{l}\phi_l (u) \phi_{n-1-l}(v)\period
\eeq
Equating this with the left hand side, we arrive at the formula (\ref{eq:decomp}).

\paragraph{Baxter-like equations}Owing to the following property of the cosh kernel, 
\beq \label{eq:coshid}
\frac{1}{\cosh \left( u+\frac{i(\pi- \epsilon)}{2}\right)}+\frac{1}{\cosh \left( u-\frac{i(\pi- \epsilon)}{2}\right)} =2\pi \delta(u) \comma
\eeq
the representation (\ref{eq:recphi}) implies a system of Baxter-like functional relations (see appendix \ref{app:technical} for more details),
\beq
\begin{aligned}
P_{\star}^{++}+P_{\star}^{--} &= 2\pi \delta_{\star} \, v \, z \, Q_{\star}   \\
Q_{\star}^{++}+Q_{\star}^{--} &= 2\pi \delta_{\star} \, v \, z \, P_{\star} \comma  \label{eq:baxterPQmt}
\end{aligned}
\eeq
with $\delta_{s}=1$, $\delta_{+}=2$ and $v =\left(1+e^{\epsilon}\right)^{-1/2}\left(1+e^{\epsilon^{++}} \right)^{-1/2}$. The functions $P_{\star}$ and $Q_{\star}$ are built out of the $\phi_{\star \,  j}$ as
\beq \label{eq:PQmt}
P_{\star}(u) \equiv E_{\star}(u)\sum_{j=0}^{\infty} z^{2j+1} \phi_{\star \, 2j+1}(u)\comma \quad Q_{\star}(u) \equiv  E_{\star}(u)\sum_{j=0}^{\infty} z^{2j} \phi_{\star \, 2j}(u) \period
\eeq
For notational purposes, we also use the superscripts $\pm$ to denote shifts by $\pm i \pi/4$ in the argument of a generic function $f$: 
\beq f^{\pm} \equiv f(u+ i\pi/4)\comma \quad f^{\pm\pm} \equiv f(u\pm i\pi/2)\quad {\rm{ and }}\quad f^{\overbrace{\pm \dots \pm}^{a}} \equiv f(u\pm i a \pi/4) \period
\eeq

\paragraph{Functional equations} We would like now to convert the Baxter-like equations (\ref{eq:baxterPQmt}) into some functional equations for the kernels $K_{\star}$. It proves useful to split them into  their even and odd components defined as
\beq
\hat{R}_{{\rm{o}} \, \star} \equiv \hat{K}_{\star}(I-z^2 \hat{K}_{\star}^2)^{-1}\,, \quad \hat{R}_{\rm{e} \, \star} \equiv \hat{K}_{\star}^2(I-z^2 \hat{K}_{\star}^2)^{-1}
\eeq
so that for example,
\beq
\hat{R}_{\rm{e}\, \star} \cdot f(u)= \sum_{k=1}^{\infty} \int\,dv\, z^{2k} K_{\star}^{\ast 2k}(u,v) f(v) \equiv \int \,dv\, R_{\rm{e}\,\star}(u,v) f(v)
\eeq
and similarly for $R_{\rm{o}\, \star}$. Here and in what follows we omit the $z$ dependence of $R_{{\rm{o}}\,\star}$ and $R_{{\rm{e}}\,\star}$ in order to make the notation simpler. Finally, we will be interested in computing the trace of the $j$-th convolution of the kernel (see expression (\ref{eq:detexpand})) for which we conveniently define
\beq
R_{{\rm{e}} \, \star}(u)\equiv R_{{\rm{e}} \,\star}(u,u)\quad\quad {\rm{and}} \quad\quad R_{{\rm{o}}\, \star}(u)\equiv R_{{\rm{o}}\,\star} (u,u) \period
\eeq
Owing to the lemma \eqref{eq:decomp}, the kernels have a simple form when written in terms of the Baxter-like functions $Q_{\star}$ and $P_{\star}$,
\beq \label{eq:RandPQ}
\begin{aligned}
& R_{{\rm{o}} \, s}(u,v)= \frac{Q_s(u)Q_{s}(v) -P_{s}(u)P_s(v)}{e^{2u}+ e^{2v}} \quad &&R_{{\rm{e}}\, s}(u,v)= \frac{Q_s(u)P_s(v) -Q_s(v)P_s(u)}{e^{2u}-e^{2v}}  \\
& R_{{\rm{o}}\,+}(u,v)= \frac{Q_{+}(u)Q_{+}(v) -P_{+}(u)P_{+}(v)}{\cosh (2u)+ \cosh(2v)}\ \quad && R_{{\rm{e}}\,+}(u,v)= \frac{Q_{+}(u)P_{+}(v) -Q_{+}(v)P_{+}(u)}{\cosh (2u) - \cosh{(2v)}} \period
\end{aligned}
\eeq
We would like now to use~(\ref{eq:baxterPQmt})  to find functional relations involving $R_{{\rm{e}} \, \star}(u)$ and $R_{{\rm{o}} \, \star}(u)$. It turns out that one needs an additional auxiliary function $\eta_{\star}$ built out of a combination of $P_{\star}$ and $Q_{\star}$
in order to close the system of equations. We refer the reader to the appendix~\ref{app:technical} where the full details are provided. Here we simply quote the final result,
\beq
\log(1+\eta_{\star}^2) = \log(1+e^{\epsilon^{+}})+\log(1+e^{\epsilon^{-}})+\log \tilde{R}_{{\rm{e}}\,\star}^{+}+\log \tilde{R}_{{\rm{e}}\,\star}^{-}\comma  \label{eq:fun1mt}
\eeq
\beq
\frac{2i\, \eta'_{\star}}{\eta_{\star}^2+1} = 2 i \arctan(\eta_{\star})' = \frac{\tilde{R}^{+}_{{\rm{o}}\,\star}}{\tilde{R}^{+}_{{\rm{e}}\,\star}}-\frac{\tilde{R}^{-}_{{\rm{o}}\,\star}}{\tilde{R}^{-}_{{\rm{e}}\,\star}} \comma  \label{eq:fun2mt}
\eeq
\beq
\eta^{+}_{\star} +\eta^{-}_{\star} = 2 \pi \delta_{\star}\left( \frac{\cosh(u)}{\sinh(u)} \right)^{\delta_{\star}-1}\, \tilde{R}_{{\rm{e}}\,\star}(u) \comma  \label{eq:fun3mt}
\eeq
where the tilde function $\tilde{R}_{{\rm{e}}\, \star}$ and $\tilde{R}_{{\rm{o}}\,\star}$ have a very simple relation with the  corresponding untilde ones given in the appendix~\ref{app:technical}.
These equations are the basis for the Tracy-Widom TBA to be derived in the next section.

\subsection{Inversion and Tracy-Widom TBA} 
We will now discuss how to use the system of functional equations  (\ref{eq:fun1mt}), (\ref{eq:fun2mt}) and (\ref{eq:fun3mt}),  along with certain analytic properties of the functions involved, to obtain integral equations that determine $R_{{\rm{e}}\,\star}$ and $R_{{\rm{o}}\,\star}$ uniquely. The basic idea behind the inversion of this type of functional equations is to consider them in Fourier space where the non-local relations become local as the shifts in the argument can be easily undone. However, that can only be performed provided one has some analytic control of the equations. More precisely, in order to Fourier transform them one must ensure that the transform itself converges. Besides that, to undo the shifts one must account for the singularities (if any) as the integration contour is deformed.

For the kernel $\star=s$, these conditions have already been thoroughly studied by Tracy and Widom \cite{Tracy:1995ax} and we will not redo that analysis here. The outcome is that the two sides of the functional equations are free of singularities within the strip $|\Im u|< \pi/4$ and they are $L^2$ functions. Therefore it is safe to Fourier transform them and undo the shifts in the standard way to obtain the first set of Tracy-Widom TBA equations,
\beq \label{eq:TWTBA1}
\begin{aligned}
\eta_s &= 2\int_{-\infty}^{\infty} dv \frac{R_{{\rm{e}}\,s}(v)}{\cosh(2(u-v))} \\
R_{{\rm{e}}\,s}(u) &=\frac{1}{1+e^{\epsilon(u)}} \exp\left( \frac{1}{2\pi}\int_{-\infty}^{\infty} dv \frac{\log(1+\eta_{s}^{2}(v))}{\cosh(2(u-v))} \right) \\
R_{{\rm{o}}\,s}(u) &= \frac{R_{{\rm{e}}\,s}(u)}{\pi}\int_{-\infty}^{\infty}dv \frac{\arctan (\eta_s(v))}{\cosh(2(u-v))^2} \period
\end{aligned}
\eeq

For the kernel $\star=+$, although the inversion of the equations  (\ref{eq:fun1mt}) and  (\ref{eq:fun2mt}) is analogous due to the same analytic properties, the inversion of (\ref{eq:fun3mt}) is slightly different. By definition (see appendix \ref{app:technical}), $\eta_{+}$ contains poles at $u=\pm i\pi/4$  and the RHS of equation (\ref{eq:fun3mt}) contains a pole at $u=0$. To correctly invert it we introduce an $i\epsilon$-prescription and do the Fourier transform with an integration contour parallel to the real axis at $u-i \epsilon$ with $u \in \mathbb{R}$. We can then freely shift the $\eta_{+}^{+}$ to the real axis.  We can also shift $\eta_{+}^{-}$ to the real axis, at the price of adding the contribution of half the residue of the pole at $u=- i\pi/2$ that crossed the contour along the way. That contribution can be traded by a principal value integration. After some straightforward manipulations, we arrive at the following integral equations
\beq
\begin{aligned}  \label{eq:TWTBA2}
\eta_+ &= 4\, \dashint_{-\infty}^{\infty} dv\, \frac{\coth(2 v) \, R_{{\rm{e}}\,+}(v)}{\cosh(2(u-v))} \\
R_{{\rm{e}}\,+}(u) &=\frac{\cosh(u)^2}{\cosh(2 u)\, (1+e^{\epsilon(u)})} \exp\left( \frac{1}{2\pi}\int_{-\infty}^{\infty} dv\frac{\log(1+\eta_{+}^{2}(v))}{\cosh(2(u-v))} \right) \\
R_{{\rm{o}}\,+}(u) &= \frac{2 \,R_{{\rm{e}}\,+}(u) \coth(2 u)}{\pi}\int_{-\infty}^{\infty} dv \frac{\arctan (\eta_+(v))}{\cosh(2(u-v))^2}\comma
\end{aligned}
\eeq
where in the first equation the integral is performed in the principal value sense. These equations are the main result of the paper. It is now simple to implement and very efficiently iterate them numerically. This is the subject of the next section.


\section{Numerics and comparison with CFT\label{sec:numerics}}
In this section, we will make some tests of the Tracy-Widom TBA equations. We first solve them numerically and compare the results with the brute force numerical evaluation of the Fredholm determinants. We then study the limit of small radius and make contact with some analytic results available from the boundary Liouville theory.
\subsection{Numerical evaluation of Fredholm determinants and checks}
In order to numerically implement the Tracy-Widom TBA equations (\ref{eq:TWTBA1}) and (\ref{eq:TWTBA2}), we will consider a power series expansion  in $z$  of each of the functions involved\footnote{A similar numerical routine  was described in \cite{Hatsuda:2012hm}.}
\beq
\eta_{\star} = \sum_{k=1}^{\infty}\eta_{\star}^{(k)} z^{k}\comma \quad R_{{\rm{e}}\,\star} = \sum_{k=0}^{\infty}R_{{\rm{e}}\,\star}^{(2k)} z^{2k}\comma \quad R_{{\rm{o}}\,\star} = \sum_{k=0}^{\infty}R_{{\rm{o}}\,\star}^{(2k+1)} z^{2k+1}\,.
\eeq
We then plug these expressions in (\ref{eq:TWTBA1}) and (\ref{eq:TWTBA2}) and collect powers of $z$. This allows us to solve the equations recursively: the term $\eta_{\star}^{(k)}$ is obtained from the knowledge of $ R_{{\rm{e}}\,\star}^{(2k-2)}$ and that allows us to compute $ R_{{\rm{e}}\,\star}^{(2k)}$ and $R_{{\rm{o}}\,\star}^{(2k-1)}$, provided we have first solved the standard bulk TBA which enters in the seed for this recursion through $\epsilon(u)$. The first few terms obtained in this way read as follows
\beq
\begin{aligned}
R_{{\rm{e}}\,\star}^{(0)}(u) &= \frac{1}{1+e^{\epsilon(u)}} \left(\frac{\cosh(u)^2}{\cosh(2u)} \right)^{\delta_{\star}-1}
\\
\eta_{\star}^{(1)}&= 2\, \delta_{\star}\, \dashint dv \frac{(\coth(2v))^{\delta_{\star}-1}}{\cosh(2(u-v))}\,  R_{{\rm{e}}\,\star}^{(0)}(v)  
\\
R_{{\rm{e}}\,\star}^{(2)}(u) & =\frac{1}{\pi} \int dv \frac{(\eta_{\star}^{(1)}(v))^2}{\cosh(2(u-v))}
\\
R_{{\rm{o}}\,\star}^{(1)}(u) & =\frac{2\,  R_{{\rm{e}}\,\star}^{(0)}(u) (\coth(2u))^{\delta_{\star}-1}}{\pi} \int dv \frac{\eta_{\star}^{(1)}(v) \,  }{\cosh(2(u-v))^2} \comma
\end{aligned}
\eeq
where the principal value integral in the second formula only applies to $\star=+$ and becomes an ordinary integral for $\star=s$.
\begin{table}[t]
\centering
\begin{tabular}{cccc} \toprule
                         & \multicolumn{1}{c}{$n$} & \multicolumn{1}{c}{{\rm{Tracy-Widom TBA} }} & \multicolumn{1}{c}{{\rm{Numerical integration} }} \\
                         \midrule
\multirow{3}{*}{${\rm{tr}}\, K_{+}^{\ast n} $} & 2                     &     {\bf{0.028220819491}}0                            &            {\bf{0.02822081949115}}                  \\
                         & 3                     &      {\bf{0.004671352476}}9                            &       {\bf{0.0046713524}}770                                 \\
                         & 4                     &   {\bf{0.000779870829}}8                       &  {\bf{0.00077987}}11                              \\
                         \midrule
\multirow{3}{*}{${\rm{tr}}\, K_{s}^{\ast n} $} & 2                     & {\bf{0.03080536812851}}2                            &          {\bf{0.03080536812851}}2                             \\
                         & 3                     &   {\bf{0.00479994178961}}5                        &             {\bf{0.004799941}}78                     \\
                         & 4                     &   {\bf{0.00078634624811}}5                         &          {\bf{0.00078634}}60              
                         \\ \bottomrule               
\end{tabular}
\caption{We have computed the first  few powers of the two kernels using both the Tracy-Widom TBA  and  by performing the multiple integrations. While through the TBA it is straightforward to generate arbitrary high powers of the kernel with very high precision, we find the direct integration to be feasible to a certain precision for the first few powers but  much more demanding beyond that. We show here the agreement between both approaches to some precision for the choice $m L =1$. The digits in bold are significant (on the TBA side one can easily increase the accuracy).}    \label{tab:data1}
\end{table}
The higher order terms are lengthier but equally straightforward to be obtained. 
We can then compute each term in the series expansion of the Fredholm determinant (\ref{eq:detexpand}) by the formulas
\beq
\begin{aligned}
{\rm{tr}}\, K_{\star}^{\ast 2n+1} \equiv \int du \, K_{\star}^{\ast 2 n+1} (u,u) &= \frac{1}{\pi^{2n+1}} \int du \, R_{{\rm{e}}\,\star}^{(2n)}(u) 
\\
{\rm{tr}} \,K_{\star}^{\ast 2n} \equiv \int du \, K_{\star}^{\ast 2 n} (u,u) &= \frac{1}{\pi^{2n}} \int du \, R_{{\rm{o}}\,\star}^{(2n-1)}(u) \period
\end{aligned}
\eeq
With this method, we can truncate the sum in (\ref{eq:detexpand}) for arbitrarily many terms and that determines the Fredholm determinant to the desired high precision. As a test, we can compare the formulas for the $n$-th term of the expansion with the direct computation of the corresponding $n$-th fold integral. 
The results showing agreement between both approaches are summarized in table \ref{tab:data1}.
\begin{table}[t]
\centering 
\begin{tabular}{cccc} \toprule
$m L$    & 1 & 2 & 4 \\ \midrule
$g$ & 0.9124146589  &0.9609414170   &  0.9937167650  \\
\bottomrule
\end{tabular}
\caption{A few examples of  the $g$-function for a choice of the boundary parameters ($\eta=\pi/2$ and $\vartheta=0$) and for different values of the sinh-Gordon mass and radius. The sum (\ref{eq:detexpand}) was truncated to 10 terms. We only show the significant digits, but one can straightforwardly increase the accuracy.}    \label{tab:data2}
\end{table}

We can finally assemble together the Fredholm determinant with the remaining ingredients of the $g$-function as in (\ref{eq:finalgfunction}), namely the expression of the reflection and bulk scattering matrices, and get explicit numbers out for a given choice of the sinh-Gordon radius $L$ and mass $m$. A few examples are given in table \ref{tab:data2}.

%

\subsection{Boundary sinh-Gordon and Liouville conformal data}
Having at hand all the tools to determine the $g$-function to very high precision, we can proceed to make a comparison with the expected CFT behavior in the limit of $R\rightarrow 0$.

\subsubsection{sinh-Gordon in the UV}
Let us recall the ``open string''  channel formulation of the (boundary) sinh-Gordon. The bulk theory is defined by the Lagrangian density (\ref{eq:SGlag}) integrated over a strip of width $R$. To this Lagrangian we add boundary interacting terms on the left and right boundaries which we choose to be the same for simplicity. Each integrable boundary is furthermore characterized by the two complex parameters $\eta$ and $\vartheta$ encoded in the reflection matrix ({\ref{eq:refmatrix}}). Again for simplicity, we will set $\vartheta=0$ such that the boundary problem is fully characterized by a single parameter. In summary, we end up with the following Lagrangian
\beq
L =  \int_{0}^{R} dx \left(\frac{1}{4\pi} (\partial \phi)^2 + 2\mu \cosh (2b\phi)\right) + 2\mu_{B}\left(  \cosh (b\,  \phi)|_{x=0}+ \cosh (b \, \phi)|_{x=R} \right) \comma
\eeq
where the cosmological constant $\mu_{B}$ is related to the parameter $\eta$ by \cite{Zamunpub,Bajnok:2001ug}
\beq \label{eq:bcosmo}
\mu_{B}=\sqrt{\frac{\mu}{\sin (\pi b^2)}} \cosh( b^2 \eta)\comma
\eeq
and we will often use instead the parameter $s$  defined by $ \pi s= b\, \eta $,  as $s$ is self-dual under the duality transformation $b \rightarrow 1/b$.

As described in \cite{Zamolodchikov:1995aa}, owing to the scaling properties of the operator $\cosh (2 b\phi)$ we can rescale the width of the strip to $2\pi$, and incorporate the $R$ dependence on the cosmological constants. We obtain
\beq
\begin{aligned}
L &=  \int_{0}^{2\pi} dx \left(\frac{1}{4\pi} (\partial \phi)^2 + \mu \left( \frac{R}{2\pi}\right)^{2+2 b^2}( e^{2b\phi}+e^{-2b\phi})\right) \\
&+ \mu_{B}\left( \frac{R}{2\pi}\right)^{1+ b^2} \left( ( e^{b\phi}+e^{-b\phi})|_{x=0}+( e^{b\phi}+e^{-b\phi})|_{x=R} \right) \period
\end{aligned}
\eeq
In the ultraviolet limit $R\rightarrow 0$, we can neglect one of the exponentials of the interacting term so that the theory gets reduced to the boundary Liouville. Switching to the ``closed string'' channel, it is then natural to expect that the $g$-function is related to the overlap of a bulk and boundary states in Liouville theory. The goal of this section is to make this connection more precise.

Let us describe the bulk and boundary states in both theories using the ``closed string''  channel formulation.
A more intuitive way of thinking of the space of states in the (bulk) Liouville theory consists in studying the quantum mechanical model of the zero mode of the Liouville field $\phi(x)$. This is obtained by considering a field configuration which is constant over space,
\beq \label{eq:zmode}
\phi(t,x) =\phi_0(t)\,.
\eeq
In this approximation, the states are characterized by the eigenvalue $P$ of the canonical momentum conjugate to the zero mode, which is a continuous parameter (often called \textit{Liouville momentum}), and discrete modes.

If we furthermore restrict ourselves to the classical limit where $b\rightarrow 0$ and take the momentum $P$ to be also of order $b$, then the nontrivial dynamics gets restricted to the zero mode $\phi_0$ with the additional discrete modes being negligible (this is the  so-called \textit{minisuperspace} approximation). 

\begin{figure}[t]
\centering
\includegraphics[width=14cm]{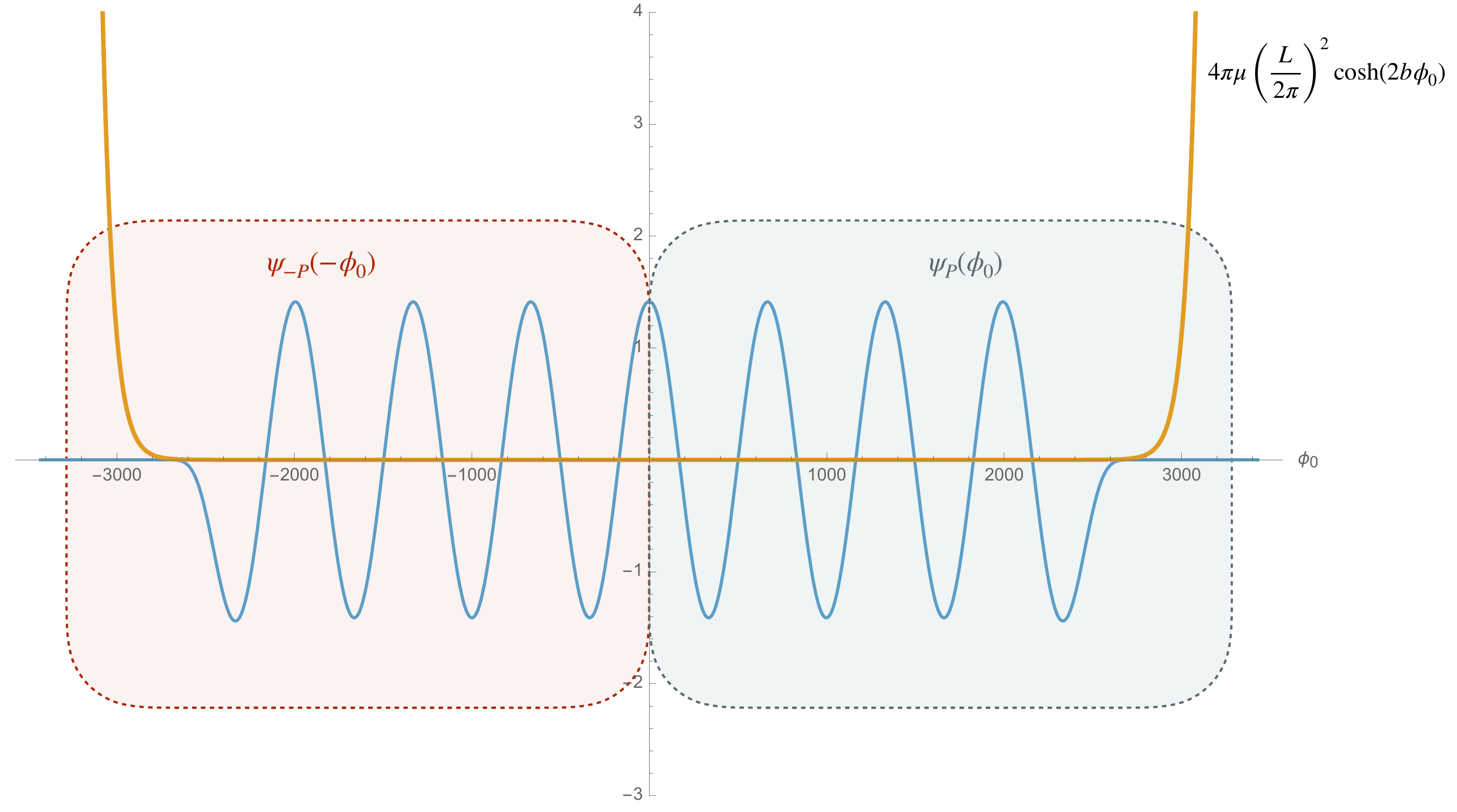}
\caption{The sinh-Gordon potential and the corresponding wave-function. In the two colored regions, we can approximate the wave-function by the corresponding Liouville one. In the middle region, they reduce to plane waves and requiring their compatibility enforces the quantization condition (\ref{eq:quantization}).} \label{potential}
\end{figure}

We can also use the same approximation for the bulk sinh-Gordon theory. We define $\phi_0$ analogously to (\ref{eq:zmode}) and the corresponding conjugate momentum $P(L)$, which is now a function of the radius $L$.  In the UV limit, the ground-state energy $E(L) = -\pi c_{{\rm{eff}}}/ 6 L$ approaches the CFT behaviour where the effective central charge becomes the Liouville central charge, $c_{{\rm{eff}}} \rightarrow c_{L} = 1-24 P^2$ with $P$ being the Liouville momentum. In the limit $L\rightarrow 0$ we expect the effective central charge to have the same dependence on the momentum up to corrections in $L$\cite{Zamolodchikov:1995aa},
\beq
c_{{\rm{eff}}} = 1-24\, P(L)^2 +\mathcal{O}(L^2)\comma
\eeq
where $P(0)$ coincides with the Liouville momentum. From now on, we will omit its $L$ dependence and simply denote it by $P$.
In this minisuperspace approximation, the bulk ground-state is described by a wave-function $\psi_{P}(\phi_0)$  satisfying the Sch\"{o}dinger equation which in this case takes up the form of an hyperbolic Mathieu equation, 
\beq  \label{eq:mathieu}
\left(-\frac{1}{2}\frac{d^2}{d\phi_{0}^2} +4 \pi \mu \left(\frac{L}{2\pi}\right)^{2} \cosh(2 b\phi_0) -2 P^2 \right)\Psi_{P}(\phi_0) = 0\,.
\eeq
As we focus on the region for $\phi_{0}>0$, the left wall of the potential localized at $2 b \phi_0 \sim \log(\mu \left(\frac{L}{2\pi}\right)^2)$ has a small effect on the solution in the CFT limit and we expect it to converge to the Liouville wave-function, see figure \ref{potential},
\beq \label{eq:Besselright}
\Psi_{P}(\phi_0) \simeq \frac{2}{\Gamma(-2i P/b)} \left(\frac{ \pi \mu L^2 }{4\pi^2 b^2}\right)^{-i P/b} K_{2 i P/b} \left(\frac{\sqrt{ \pi \mu }L }{2\pi b}e^{b\phi_0}\right)\equiv \psi_P(\phi_0)\comma
\eeq
where $K_{2i P/b}$ is a modified Bessel function.
On the other hand, for $\phi_0<0$ we expect to have the mirror solution, obtained from the above by $P\rightarrow -P$ and $\phi_{0}\rightarrow -\phi_0$. Around the middle region, $\log(\mu \left(\frac{L}{2\pi}\right)^2) \ll2b \phi_0\ll-\log(\mu \left(\frac{L}{2\pi}\right)^2)$, both potential walls are far apart and one expects the wave-functions to be well approximated by a plane wave. Taking the plane-wave limit of the region $\phi_0>0$ (which amounts to take $\phi_0 \rightarrow -\infty$ in (\ref{eq:Besselright})) we obtain,
\beq
\psi_{P}(\phi_0)  \simeq e^{2 i P \phi_0}+ S_{{\rm{c}}}(P)e^{-2 i P \phi_0} \comma
\eeq
with the (classical) reflection matrix $S_{{\rm{cl}}}(P)$ given by
\beq
S_{{\rm{cl}}}(P) = -  \left(\frac{ \pi \mu L^2 }{4\pi^2 b^2}\right)^{-2i P/b} \frac{\Gamma(1+2 i P/b)}{\Gamma(1-2 i P/b)} \period
\eeq
By considering the same plane-wave limit in the region $\phi_0<0$ and requiring compatibility of both wave-functions, we obtain the \textit{quantization condition} of the momentum $P$,
\beq \label{eq:quantization}
S_{{\rm{cl}}}(P)^2 \left( \frac{L}{2\pi}\right)^{-8 i P/b }=1 \period
\eeq
This is the classical version ($b\rightarrow 0$) of the quantization condition described in \cite{Zamolodchikov:1995aa}. We will use it to define the momentum as a function of $L$. In particular, from this relation, we get that for the ground-state, the momentum $P$ goes to zero as $L\rightarrow 0$,
\beq 
P \sim  -\frac{1}{\log(L/2\pi )} +\mathcal{O}\left(\frac{1}{\log(L/2\pi)^2} \right)\,.
\eeq
We then conclude that the ground state of sinh-Gordon gets mapped in its CFT limit into a state of zero-momentum in the Liouville theory.

\subsection{Liouville boundary data}
The object of interest for comparison with sinh-Gordon $g$-function is the bulk one-point function of the boundary Liouville theory whose formula was bootstrapped in \cite{Fateev:2000ik}. In terms of the Liouville momentum $P$ and the boundary parameter $s$ (see discussion around (\ref{eq:bcosmo})), its expression reads
\beq
\langle B_{s} | \psi_{P}\rangle_{{\rm{L}}} = \left(\pi \mu \gamma(b^2) \right)^{-i P/b} \Gamma(1+2 i b P)\,\Gamma(1+2 i  P/b) \frac{\cos(2 \pi P s)}{i P}\comma
\eeq
where the bulk states are normalized according to
\beq \label{eq:normal}
\langle \psi_{P_1}| \psi_{P_2}\rangle = \pi\, \delta(P_1-P_2)\,.
\eeq
An immediate problem with this expression for comparison  with the $g$-function is the existence of the IR singularity manifest in the pole at $P=0$, which is precisely the point we are interested in. We would like to find the correct prescription to subtract this pole and obtain a finite result. This is generally a nontrivial problem but we can obtain some insight by studying it within the minisuperspace approximation and eventually guess the full quantum result from it.

Classically, the one-point function $\langle B_{s} | v_{P}\rangle^{{\rm{cl}}} $ boils down to the overlap between the bulk  and  boundary wave-functions. The bulk wave-function is a solution of the Schr\"{o}dinger equation (\ref{eq:mathieu}) while the boundary one is related to the boundary Lagrangian by
\beq
\varphi_{B}(\phi_0) = \exp\left( -2\, \mu_B L \cosh(b \phi_0 )\right)
\eeq
and the classical sinh-Gordon one-point function is simply
\beq
\langle B_{s} | \Psi_{P}\rangle_{\rm{shG}}^{{\rm{cl}}} = \int_{-\infty}^{\infty} d\phi_0\,\Psi_{P}(\phi_0) \varphi_{B}(\phi_0) \comma
\eeq
where the superscript `${\rm{cl}}$' emphasises that this is the classical result.
Rather than attempting at computing this integral, let us relate it to $\langle B_{s} | \psi_{P}\rangle_{{\rm{L}}}^{{\rm{cl}}}$ in this limit. For that we split the integration in two regions, $\phi_0<0$ and $\phi_0>0$,
\beq
\langle B_{s} | \Psi_{P}\rangle_{\rm{shG}}^{{\rm{cl}}}  =  \int_{-\infty}^{0} d\phi_0\,\Psi_{P}(\phi_0) \varphi_{B}(\phi_0) + \int_{0}^{\infty} d\phi_0\,\Psi_{P}(\phi_0) \varphi_{B}(\phi_0)\period
\eeq
In each of these regions and in the limit $L\rightarrow 0$, the integrand is well approximated by the corresponding Liouville counterparts,
\beq
\langle B_{s} | \Psi_{P}\rangle_{\rm{shG}}^{{\rm{cl}}}  \simeq \int_{-\infty}^{0} d\phi_0\,\psi_{-P}(-\phi_0)\, e^{ -\, \mu_B L \exp(-b \phi_0 )}+  \int_{0}^{\infty} d\phi_0\,\psi_{P}(\phi_0)\, e^{ -\, \mu_B L \exp(b \phi_0 )}\period
\eeq
We can complete the integration regions over the full real axis and compensate with the appropriate subtractions. We obtain
\beq
\begin{aligned}
\langle B_{s} | \Psi_{P}\rangle_{\rm{shG}}^{{\rm{cl}}}  &\simeq \langle B_{s} | \psi_{-P}\rangle_{\rm{L}}^{{\rm{cl}}}  + \langle B_{s} | \psi_{P}\rangle_{\rm{L}}^{{\rm{cl}}}   \\
&-\frac{i}{2P}\left(S_{{\rm{cl}}}(P)-\frac{1}{S_{{\rm{cl}}}(P)}\right)+\mathcal{O}(L)
\end{aligned}
\eeq
where in the second line we have computed the pole subtractions by approximating the corresponding wave-functions by plane-waves. Interestingly, the second line is independent of the boundary parameter $s$ and as such we rather consider the one-point function relative to a reference boundary state with parameter $s_0$, 
\beq
U_{s_1,s_0}(P) \equiv \frac{\langle B_{s_1} | \Psi_{P}\rangle_{\rm{shG}}-\langle B_{s_0} | \Psi_{P}\rangle_{\rm{shG}}}{\sqrt{\langle \Psi_{P}| \Psi_{P}\rangle }} \,.
\eeq
The above classical considerations suggest that the full quantum $U_{s_1,s_0}(P)$ is given in terms of the Liouville one-point function as
\beq \label{eq:tomatch}
U_{s_1,s_0}(P) =  \frac{1}{\sqrt{\pi}}\left( \langle B_{s_1} | \psi_{-P}\rangle_{\rm{L}}  + \langle B_{s_1} | \psi_{P}\rangle_{\rm{L}}\right)- \frac{1}{\sqrt{\pi}}\left( \langle B_{s_0} | \psi_{-P}\rangle_{\rm{L}} + \langle B_{s_0} | \psi_{P}\rangle_{\rm{L}} \right) \,.
\eeq
where we have accounted for the normalization (\ref{eq:normal})\footnote{The momentum $P$ got discretized so that we have considered a normalization with Kronecker delta. Drawing an analogy with the form factors of local operators in integrable models, we might expect that in the conversion between the discrete and continuum normalizations, the analog of the \textit{Gaudin norm} might play a role. Such norm would be obtained from the momentum quantization condition in analogy to the Bethe equations. We have verified that the inclusion of such normalization factor makes the agreement worse. It would be important to further investigate this question.}.
\begin{figure}[t]
\centering
\includegraphics[width=17cm]{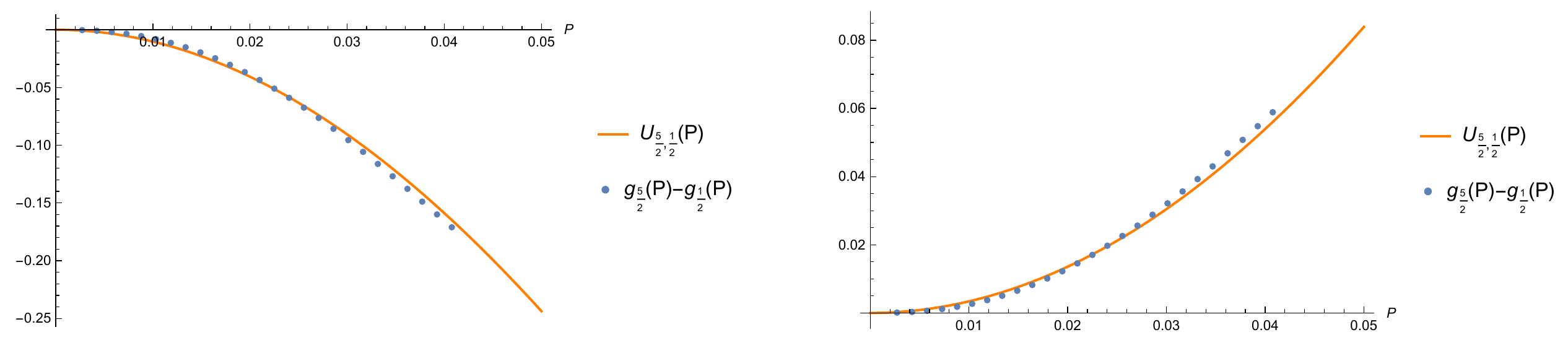}
\caption{Comparison of the difference of $g$-functions for different values of $s$ and the combination (\ref{eq:tomatch}) of the bulk Liouville one-point functions.} \label{graphs}
\end{figure}
We would like to compare this quantity with the $g$-function, or rather, the difference of $g$-functions for two different boundary parameters $s_1$ and $s_0$. The $g$-function can be plotted as an approximate function of $P$, where $P$ is determined  by the quantum version of (\ref{eq:quantization}), see appendix \ref{app:liouville} for details. We expect that for very small radius $L$ or, equivalently very small $P$, the relation (\ref{eq:quantization}) becomes more accurate and $P$ defined as such approaches the Liouville momentum. The results are shown in figure \ref{graphs}.  We have made the comparison for many different choices of the boundary parameters and here we display two such choices for illustration. In all cases, it became apparent that for small momentum there is a qualitative agreement between both quantities. We should emphasize that this match is, nevertheless, not expected to be optimal away from $P=0$ as the relation  (\ref{eq:quantization}) is approximate and, adding to this, the small radius limit is numerically harder to achieve. It would be interesting to better understand how to exactly solve the Tracy-Widom TBA in the UV limit along the lines of \cite{Bajnok:2007ep} for the sinh-Gordon UV central charge and perform a more analytic comparison with Liouville.

\section{Separation of variables and $g$-function\label{sec:sov}}
In this section, we present multiple integrals which we conjecture to describe a universal part of the $g$-functions in the sinh-Gordon theory. It is a natural generalization of the formula for the finite-volume one-point function by Lukyanov \cite{Lukyanov:2000jp}, which is based on Skylanin's separation of variables (SoV) \cite{Sklyanin:1995bm}. We first explain basic properties of Lukyanov's integral formula to motivate our proposal. We then present our conjecture and discuss its properties. Unlike the other sections, here we consider the sinh-Gordon theory at generic coupling $b$. 
\subsection{Q-Function and Lukyanov's formula}
Lukyanov's formula is written in terms of the $Q$-function, which was first discussed by Zamolodchikov in \cite{Zamolodchikov:2000kt}. It is related to the pseudo-energy of TBA as\footnote{In this section, we basically follow the notations in \cite{Kostov:2019sgu} (and partly \cite{Negro:2013wga}). The only difference is that here $f^{\pm}$ mean the shift of the arguments by $\pm i\pi/4$ while in \cite{Kostov:2019sgu} they mean the shift by $\pm i\pi/2$.}
\beq
1+e^{-\epsilon (u)}=Q^{++}(u)Q^{--}(u)\comma
\eeq
and satisfies the functional identity
\beq\label{eq:QQrelation}
Q^{++}(u)Q^{--}(u)=1+Q^{[2a]}(u)Q^{[-2a]}(u)\comma
\eeq
with $a\equiv (1-b^2)/(1+b^2)$ and $Q^{[\pm 2a]}(u)\equiv Q(u\pm \frac{ia\pi}{2})$. Another important quantity is the $T$-function which is defined by\footnote{Note that \eqref{eq:BaxterTQ} can also be viewed as the Baxter TQ-relation.}
\beq\label{eq:BaxterTQ}
T(u)\equiv \frac{Q^{[2(1-a)]}(u)+Q^{[-2(1-a)]}(u)}{Q(u)}\period
\eeq
Using \eqref{eq:QQrelation}, one can show the periodicity of the $T$-function
\beq\label{eq:Tperiodicity}
T^{[2(1+a)]}(u)=T(u)\period
\eeq
Thanks for the duality of the sinh-Gordon theory $b\leftrightarrow b^{-1}$, one can define a dual $T$-function,
\beq
\tilde{T}(u)\equiv \frac{Q^{[2(1+a)]}(u)+Q^{[-2(1+a)]}(u)}{Q(u)}\comma
\eeq
which satisfies the dual periodicity
\beq
\tilde{T}^{[2(1-a)]}(u)=\tilde{T}(u)\period
\eeq
These functional equations were originally derived for the ground state but they are expected to hold also for excited states \cite{Zamolodchikov:2000kt,Negro:2013wga}.

In \cite{Lukyanov:2000jp}, Lukyanov wrote down a multiple-integral formula for the one-point function in the finite volume. Specializing it to the identity operator, we obtain the following expression for the norm of the state
\beq
\begin{aligned}\label{eq:lukyanov}
&\langle \Omega |\Omega\rangle =\lim_{N\to \infty}\mathcal{I}_N\comma\\
&\mathcal{I}_N\equiv \frac{1}{(2N+1)!}\int_{-\infty}^{\infty}\prod_{k=-N}^{N}\frac{d\theta_k \left(Q(\theta_k)\right)^2}{2\pi}\prod_{-N\leq j<k\leq N} \Delta (\theta_j,\theta_k)\comma
\end{aligned}
\eeq
with
\beq
\begin{aligned}
\Delta (\theta_j,\theta_k)\equiv \left(2\sinh \nu (\theta_j-\theta_k)\right)\left(2\sinh \tilde{\nu} (\theta_j-\theta_k)\right)\comma\\
\nu\equiv \frac{2}{1+a}=1+b^2\comma\qquad \tilde{\nu}\equiv\frac{2}{1-a}=1+b^{-2}\period 
\end{aligned}
\eeq

Using the Vandermonde determinant formula, 
\beq
\prod_{-N\leq j<k\leq N} \left(2\sinh \nu (\theta_j-\theta_k)\right) =\det \left(e^{2 \nu j \theta_k}\right)_{-N\leq j,k\leq N}\comma
\eeq
this can be re-expressed as
\beq
\langle \Omega |\Omega \rangle =\lim_{N\to \infty}\frac{1}{(2N+1)!}\int_{-\infty}^{\infty}\left(\prod_{k=-N}^{N}\frac{d\theta_k \left(Q(\theta_k)\right)^2}{2\pi} \right) \det \left(e^{2 \nu j \theta_k}\right)\det \left(e^{2 \tilde{\nu} j \theta_k}\right)\period
\eeq 
Next we expand one of the determinants into a sum over permutations. Owing to the antisymmetry of the rest of the integrand, all the permutations give the same answer. We can thus pick the simplest one $\sigma_j =j$ and multiply the overall factor $(2N+1)!$. After doing so, we expand the other determinant to get
\beq
\langle \Omega |\Omega\rangle=\lim_{N\to \infty}\sum_{\sigma\in S_{2N+1}}(-1)^{|\sigma|}\int^{\infty}_{-\infty}\prod_{k=-N}^{N}\frac{d\theta_k \left(Q(\theta_k)\right)^2}{2\pi}e^{2\left(\nu \sigma_k+\tilde{\nu}k\right)\theta_k}\period
\eeq 
One can then re-organize this into a determinant of size $2N+1$,
\beq\label{eq:bigdeterminant}
\langle \Omega |\Omega \rangle=\lim_{N\to\infty}\det \left[M_{j,k}\right]_{-N\leq j,k\leq N}\comma
\eeq
with
\beq
M_{j,k}=\int_{-\infty}^{\infty}\frac{d\theta }{2\pi}(Q (\theta))^2e^{2(\nu k+\tilde{\nu}j)\theta}\period
\eeq

Before ending this subsection, let us point out one interesting property of the formulae \eqref{eq:lukyanov} and \eqref{eq:bigdeterminant} which was not discussed in the literature. For this purpose, we first conjecture that \eqref{eq:lukyanov} and \eqref{eq:bigdeterminant} can be generalized to the overlap between arbitrary excited states $\langle \Psi_1 |\Psi_2\rangle$ by the replacement
\beq
\left(Q(u)\right)^2 \mapsto Q_1 (u)Q_2(u)\comma
\eeq
where $Q_{1,2}$ denotes the $Q$-function for the state $\Psi_{1,2}$. We should emphasize that this is purely a conjecture unlike \eqref{eq:lukyanov}, which was derived from the lattice discretization. Nevertheless it reproduces the salient feature of the overlap, namely the orthogonality of different eigenstates. To see this, we consider the $T$-functions associated to these states
\beq\label{eq:BaxterT12}
T_{1,2}\equiv \frac{Q_{1,2}^{[2(1-a)]}(u)+Q_{1,2}^{[-2(1-a)]}(u)}{Q_{1,2}(u)}\period
\eeq
Since they satisfy the periodicity condition \eqref{eq:Tperiodicity}, one can expand them as
\beq
T_{1,2}(\theta)=\sum_{n=-\infty}^{\infty} a_n^{(1,2)}e^{2\nu n \theta}\period
\eeq
Let us next consider a vector
\beq
\vec{a}\equiv (\ldots, a_{-1}^{(1)}-a_{-1}^{(2)},a_{0}^{(1)}-a_{0}^{(2)},a_{1}^{(1)}-a_{1}^{(2)},\ldots)^{t}\comma
\eeq
and multiply it to the matrix
\beq
M^{(1,2)}_{j,k}\equiv \int_{-\infty}^{\infty}\frac{d\theta }{2\pi}Q_1 (\theta) Q_{2}(\theta)e^{2(\nu k+\tilde{\nu}j)\theta}\period
\eeq
We then get
\beq\label{eq:Mtimesa}
\begin{aligned}
\left(M^{(1,2)}\cdot \vec{a}\right)_{j}&=\int^{\infty}_{-\infty}\frac{d\theta}{2\pi} \left(T_1 (\theta)-T_2 (\theta)\right)Q_1 (\theta) Q_{2}(\theta)e^{2\tilde{\nu}j\theta}\\
&=\int^{\infty}_{-\infty}\frac{d\theta}{2\pi} \left(Q_2 (Q_{1}^{[2(1-a)]}+\textcolor[rgb]{1,0,0}{Q_{1}^{[-2(1-a)]}})-Q_1 (Q_{2}^{[2(1-a)]}+\textcolor[rgb]{1,0,0}{Q_{2}^{[-2(1-a)]}})\right)e^{2\tilde{\nu}j\theta}\\
&=0\period
\end{aligned}
\eeq
Here we used the Baxter equations \eqref{eq:BaxterT12} in the second line. In the third line, we shifted the arguments of the terms denoted in red by $i \pi  (1-a)/2$ using the periodicity of $e^{2\tilde{\nu}j\theta}$.
The result \eqref{eq:Mtimesa} shows that the matrix $M^{(1,2)}$ has a zero eigenvalue. From this (and \eqref{eq:bigdeterminant}), it immediately follows that the conjectured formula for the overlap $\langle \Psi_{1}|\Psi_{2}\rangle$ vanishes unless they have the same $T$-functions $T_{1}=T_2$.
\subsection{Conjecture for $g$-function\label{subsec:conjectureg}}
\paragraph{Factorization of norm}
To motivate our proposal for the $g$-function, let us first show that the determinant expression for the norm \eqref{eq:bigdeterminant} can be factorized into a product of two smaller determinants whenever the $Q$-function is parity symmetric $Q(-\theta)=Q(\theta)$. Using the parity symmetry, we rewrite the matrix element $M_{j,k}$ as
\beq
M_{j,k}=M_{-j,-k}=\int^{\infty}_{-\infty}\frac{d\theta}{2\pi}\left(Q(\theta)\right)^2 \cosh\left[2(\nu k+\tilde{\nu}j)\theta\right]\period
\eeq
We next re-order the rows and the columns of $M$ and bring it into a block-matrix form,
\beq
M=\pmatrix{c|c|c}{\text{\Large $A$}&c&\text{\Large $B$}\\\hline d^{t}&e&d^{t}\\\hline\text{\Large $B$}&c&\text{\Large $A$}}\comma
\eeq
where $A$, $B$, $c$, $d$ and $e$ read
\beq
\begin{aligned}
&A_{s,t}\equiv M_{s,t}=M_{-s,-t}\comma\qquad B_{s,t}\equiv M_{s,-t}=M_{-s,t}\comma\\
& c_s\equiv M_{s,0}=M_{-s,0}\comma\qquad d_s\equiv M_{0,s}=M_{0,-s}\comma\qquad e\equiv M_{0,0}\period
\end{aligned}
\eeq
with $1\leq s,t\leq N$. We then add and subtract rows and columns of $M$ without modifying its determinant to get
\beq
\pmatrix{c|c|c}{\text{\Large $A$}&c&\text{\Large $B$}\\\hline d^{t}&e&d^{t}\\\hline\text{\Large $B$}&c&\text{\Large $A$}} \mapsto \pmatrix{c|c|c}{\text{\Large $A-B$}&0&\text{\Large $B-A$}\\\hline d^{t}&e&d^{t}\\\hline\text{\Large $B$}&c&\text{\Large $A$}}\mapsto\pmatrix{c|c|c}{\text{\Large $A-B$}&0&\text{\Large $0$}\\\hline d^{t}&e&2d^{t}\\\hline\text{\Large $B$}&c&\text{\Large $A+B$}}\period
\eeq
This shows that $\det M$ can be factorized into
\beq\label{eq:factorizationM}
\det M = \frac{1}{2}\det M^{-} \det M^{+}\comma
\eeq
with
\beq
\begin{aligned}
&(M^{-})_{s,t}=2\int^{\infty}_{-\infty}\frac{d\theta}{2\pi}Q(\theta)Q(-\theta) \sinh (2\nu s\theta)\sinh (2\tilde{\nu} t\theta)\qquad &&(1\leq s,t \leq N)\comma\\
&(M^{+})_{s,t}=2\int^{\infty}_{-\infty}\frac{d\theta}{2\pi}Q(\theta)Q(-\theta) \cosh (2\nu s\theta)\cosh (2\tilde{\nu} t\theta)\qquad &&(0\leq s,t \leq N)\period
\end{aligned}
\eeq
Here we used the parity symmetry of the $Q$-function and rewrote $(Q(\theta))^2$ as $Q(\theta)Q(-\theta)$ for later convenience.
\paragraph{Conjecture for $g$-function} The expression \eqref{eq:factorizationM} resembles the factorization of the Gaudin norm of parity-symmetric Bethe states in integrable spin chains\footnote{See for instance \cite{deLeeuw:2017cop} and references therein.}. In that case, one of the factors gives an universal part of the overlap between the boundary state and the Bethe state, which does not depend on the details of the boundary state. 

Now, from the analogy with the Gaudin norm, we conjecture that the overlap between the boundary state and the ground state is given by $\langle B |\Omega\rangle\propto \det M^{-}$. Together with the expression for the norm $\langle \Omega |\Omega \rangle =\det M$, this leads to the following conjecture for the universal part of the $g$-function, which is given by a ratio of the Fredholm determinants \eqref{eq:finalgfunction}
\beq\label{eq:mainconjecture}
\frac{\sqrt{{\rm Det} (1-\hat{G})}}{{\rm Det} (1-\hat{G}_{+})}=\lim_{N\to \infty}\mathcal{N}\times \frac{\det M^{-}}{\sqrt{\det M}}\period
\eeq
Here $\mathcal{N}$ is a possible normalization factor which cannot be determined by this simple argument. 

Note that the full $g$-function is given by a product of \eqref{eq:mainconjecture} and the overall factor which depends on the reflection matrix (see \eqref{eq:finalgfunction}). Since the overall factor is given in terms of the pseudo-energy, it is straightforward to express it in terms of $Q$-function as
\beq
\int^{\infty}_{0} 
\frac{d\theta}{2\pi}\Theta (\theta) \log (1+e^{-\epsilon (\theta)})=\int^{\infty}_{0}\frac{d\theta}{2\pi}\Theta (\theta)\log \left(Q^{++}(\theta)Q^{--}(\theta)\right)\period
\eeq 
\paragraph{SoV-like integral for $g$-function} One can also rewrite our conjecture into multiple integrals, which can be viewed as a generalization of Lukyanov's formula. Since  $\det M$ is given by Lukyanov's formula, our task is to rewrite the denominator $\det M^{-}$ into multiple integrals. To do so, we note that $2\sinh (n\alpha \theta)$ is $2\sinh \alpha\theta$ times a polynomial of $2\cosh \alpha\theta$ of degree $n-1$:
\beq\label{eq:coshsinhidentity}
2\cosh (n\alpha \theta)=(2\sinh \alpha \theta)\times \left[\left(2\cosh \alpha \theta\right)^{n-1}+\cdots\right]\period
\eeq
Therefore, by adding and subtracting rows and columns, we can replace $M^{-}$ with the following $\overline{M}^{-}$ without changing its determinant ($\det M^{-}=\det \overline{M}^{-}$):
\beq
(\overline{M}^{-})_{s,t}=\int^{\infty}_{-\infty}\frac{d\theta \sinh (2\nu\theta)\sinh (2\tilde{\nu}\theta)}{\pi}Q(\theta)Q(-\theta) \left(2\cosh (2\nu \theta)\right)^{s-1}\left(2\cosh (2\tilde{\nu} \theta)\right)^{t-1}\period
\eeq
Now using the Vandermonde determinant formula, one can re-express $\det \overline{M}^{-}\left(\equiv \overline{\mathcal{I}}_N\right)$ as follows:
\beq\label{eq:integralMbar}
\begin{aligned}
&\overline{\mathcal{I}}_N=\frac{1}{N!}\int^{\infty}_{-\infty}\left(\prod_{k=1}^{N}\frac{d\theta_k\sinh (2\nu\theta_k)\sinh (2\tilde{\nu}\theta_k) Q(\theta_k)Q(-\theta_k)}{\pi}\right)\prod_{1\leq j,k\leq N}\overline{\Delta} (\theta_j,\theta_k)\comma
\end{aligned}
\eeq
with
\beq
\begin{aligned}
\overline{\Delta} (\theta_j,\theta_k)&\equiv \left[2\cosh (2\nu\theta_j)-2\cosh (2\nu\theta_k)\right]\left[2\cosh (2\tilde{\nu}\theta_j)-2\cosh (2\tilde{\nu}\theta_k)\right]\\
&=\left(\sinh^2 (\nu\theta_j)-\sinh^2 (\nu\theta_k)\right)\left(\sinh^2 (\tilde{\nu}\theta_j)-\sinh^2 (\tilde{\nu}\theta_k)\right)\period
\end{aligned}
\eeq
Combined with Lukyanov's formula \eqref{eq:lukyanov}, the expression \eqref{eq:integralMbar} gives the SoV-like integral representation for the ratio of Fredholm determinants,
\beq\label{eq:INIbarN}
\frac{\sqrt{{\rm Det} (1-\hat{G})}}{{\rm Det} (1-\hat{G}_{+})}=\lim_{N\to \infty}\mathcal{N}\times \frac{\overline{\mathcal{I}}_N}{\sqrt{\mathcal{I}_N}}\period
\eeq

We must admit that our proposal is incomplete since we were unable to specify\footnote{A possible way to determine the normalization factor is to start from the lattice discretization and carefully take the continuum limit (see \cite{Lukyanov:2000jp,Bytsko:2006ut,Teschner:2007ng}). In this paper, we used the SoV representation for a lattice with an odd number of sites. However, in order to discuss the overlap, it might be more appropriate to consider a lattice with an even number of sites \cite{Bytsko:2006ut}.} the (possible) normalization factor $\mathcal{N}$.  Nevertheless we think that the results and the arguments presented here are useful for several reasons: First, despite incompleteness, our proposal reproduces one of the most important properties of the overlap $\langle B|\psi \rangle$; namely the overlap is nonzero only when the state $|\psi\rangle$ is parity-symmetric. See below for the derivation. Second, in appendix R of \cite{Jiang:2019xdz}, a similar integral representation was derived for an overlap between the boundary state and the Bethe state in the XXX spin chain. There, it was derived by rewriting an exact expression for the overlap \cite{Foda:2015nfk}. However one can alternatively arrive at the same expression by following our argument as we show in appendix \ref{ap:factorization}; namely by rewriting the SoV integral for the norm \cite{Kazama:2013rya} and factorizing it into two pieces. Such a trick would be useful for guessing the integral representation for the overlaps in higher-rank spin chains, for which the SoV integrals for the norms were discussed in \cite{Cavaglia:2019pow,Gromov:2019wmz}.

\paragraph{Selection rule} One of the important properties of the integrable boundary state is that it is annihilated by the action of higher conserved charges  which are odd under the parity transformation \cite{Ghoshal:1993tm}. Owing to this property, the overlap $\langle B|\psi\rangle$ will vanish if the state $|\psi \rangle$ is {\it not} parity-symmetric. We now show that our proposal reproduces this selection rule; namely $\det M^{-}$ vanishes if the $Q$-function is not parity-symmetric, $Q(\theta)\neq Q(-\theta)$. 

The derivation is similar to the proof of orthogonality. The starting point is to consider a linear combination of $T$-functions, $T(\theta)-T(-\theta)$. If the state is not parity symmetric, this does not vanish and can be expanded as
\beq
T(\theta)-T(-\theta)=\sum_{n=1}^{\infty} t_n \sinh (2\nu n \theta)\comma
\eeq 
since it is periodic with period $i\pi(1+a)/2$ and odd under $\theta\to -\theta$. We then consider a vector
\beq
\vec{t}\equiv (t_0,t_1,\ldots)^{t}\comma
\eeq
and multiply it to $M^{-}$. This leads to
\beq
\begin{aligned}
\left(M^{-}\cdot \vec{t}\right)_{m}=&2\int^{\infty}_{-\infty}\frac{d\theta}{2\pi}\sinh (2\tilde{\nu}m\theta)(T(\theta)-T(-\theta))Q(\theta)Q(-\theta)\\
=&2\int^{\infty}_{-\infty}\frac{d\theta}{2\pi}\sinh (2\tilde{\nu}m\theta)\left[Q(-\theta)(Q^{[2(1-a)]}(\theta)+Q^{[-2(1-a)]}(\theta))\right.\\
&\left.-Q(\theta)(Q^{[2(1-a)]}(-\theta)+Q^{[-2(1-a)]}(-\theta))\right]\\
&=0\period
\end{aligned}
\eeq
This shows that $M^{-}$ has a zero eigenvalue and therefore we have $\det M^{-}=0$.

\section{Conclusion\label{sec:conclusion}}
In this paper, we reformulated the $g$-functions in the sinh-Gordon theory in terms of the TBA-like integral equation, which we called the Tracy-Widom TBA. The resulting integral equation is more efficient than the results based on the Fredholm determinants. We only performed the analysis for a specific theory but we hope our result paves the way towards finding a more efficient formalism for $g$-functions in general integrable theories. Below we mention several future directions.
\paragraph{Excited-state $g$-function} One promising direction is to study the excited-state $g$-functions discussed first in \cite{Kostov:2018dmi} and computed in a full form in \cite{Jiang:2019xdz,Jiang:2019zig}, see also \cite{Zolitoappear}. For excited states, the result is given by a generalization of the Fredholm determinant which involves both integrals and sums. It would be interesting to see whether and how our analysis carries over to such cases. Another related question is to understand the analytic continuation of the Tracy-Widom TBA. In standard TBAs, the analytic continuation allows us to compute the spectrum of the excited states \cite{Dorey:1996re}. It would thus be interesting to see if some analytic continuation of the Tracy-Widom TBA gives the excited-state $g$-function.
\paragraph{Generalization to other theories} Another immediate question is whether one can generalize the analysis of this paper to more general integrable field theories. In particular, the generalization to theories with bound states and/or internal degrees of freedom would be important since the Fredholm determinant for such theories takes an even more complicated form, and there are also conceptual points to be understood\footnote{See for instance \cite{Kostov:2019fvw}.}.  
\paragraph{Application to $\mathcal{N}=4$ SYM} Once these two generalizations are achieved, an obvious next step is to apply this formalism to $\mathcal{N}=4$ SYM in which a special class of three-point functions can be computed by the excited-state $g$-functions. Such a reformulation would allow us to perform the numerical computation more efficiently and understand how the finite-coupling $g$-function interpolates between the weak and strong coupling results.

Putting it into a broader context, it would be important to find functional/integral equations which directly compute the correlation functions of $\mathcal{N}=4$ SYM. Currently the most advanced method for computing the correlation functions from integrability is the hexagon formalism \cite{Basso:2015zoa}. Although it in principle provides a way to compute nonperturbative correlation functions, it is desirable to find an alternative since the hexagon formalism gives an infinite series of the intermediate states which is sometimes hard to evaluate. Recently there are some hints \cite{Cavaglia:2018lxi,Giombi:2018qox,Giombi:2018hsx} that the correlation functions may be reformulated as integrals of the so-called Quantum Spectral Curve \cite{Gromov:2013pga,Gromov:2014caa}---the most efficient formalism for computing the spectrum. However, even if this is achieved, it is not totally clear if such integrals are really tractable\footnote{In fact, the SoV-like integral we conjectured in this paper is very hard to evaluate in practice.} at finite coupling.  Thus it would be important to find an analogue of TBA or Quantum Spectral Curve which directly computes the correlation functions. We hope our work provides a small step towards such an ambitious goal.
  
  \paragraph{Application to $S^3$ partition function} In this paper, we generalized the Tracy-Widom TBA in \cite{Tracy:1995ax,Okuyama:2015pzt} to the kernel given by \eqref{eq:kernelplusEandM}. The same kernel shows up also in the computation of the $S^3$ partition functions of superconformal Chern-Simons theory with orthosymplectic gauge groups \cite{Moriyama:2015asx,Okuyama:2015auc,Honda:2015rbb,Moriyama:2016xin,Okuyama:2016xke}. It should therefore be possible to derive the Tracy-Widom TBA for those theories and use it to analyze nonperturbative corrections. 
  
  \paragraph{Separation of variables} Another important future problem is to sharpen our conjecture for the SoV-like integral (and correct it if needed) by starting from the lattice discretization \cite{Lukyanov:2000jp,Bytsko:2006ut,Teschner:2007ng} and taking the continuum limit. It would also be interesting to apply our argument to higher-rank and/or non-compact\footnote{The Gaudin-like determinant formula for the overlap in the SL(2) spin chain was conjectured in \cite{Jiang:2019xdz} and proven recently in \cite{Jiang:2020sdw}. In appendix \ref{ap:factorization}, we present a conjecture for the SoV representation of such overlaps.} spin chains and guess the SoV representation for the overlap between the boundary state and the Bethe state. 
  
\paragraph{Other directions} Our result for the $g$-function has a ``double-layer'' structure. To compute it, we first solve the standard TBA and then plug the result into the Tracy-Widom TBA. This might appear slightly unusual, but a similar double-layer structure shows up in the analysis of the wall crossing of the 2d-4d coupled system \cite{Gaiotto:2011tf} (see for instance (5.21)-(5.24)). It would be interesting\footnote{It would be fascinating if one can directly relate the two by considering some supersymmetric integrable theories with boundaries.} to compare the two and understand the similarities and differences. A somewhat related question is whether our result has any implication to the ODE/IM correspondence \cite{Dorey:1998pt}\footnote{See also the review \cite{Dorey:2007zx} and more recent developments \cite{Ito:2018eon,Ito:2019llq}.}. The ODE/IM correspondence allows us to compute the spectrum of certain Schr\"{o}dinger equations from a standard TBA. It would be interesting to find counterparts of the $g$-function\footnote{A natural guess is a wave function itself.} and the Tracy-Widom TBA in that context.
\subsection*{Acknowledgement} 
We thank Alba Grassi and Ivan Kostov for helpful dicussions on the Fredholm determinant and TBA. We also thank Stefano Negro, Leonardo Rastelli, Fedor Smirnov and Alexander Zamolodchikov for discussions and comments. We thank Zoltan Bajnok and Yasuyuki Hatsuda for comments on the draft. J.C.~is supported by a Simons Collaboration grant and S.K.~is supported by DOE grant number DE-SC0009988.
\appendix
\section{Derivation of the functional equations} \label{app:technical}
In this appendix we provide the technical details of our derivation of the functional equations for the Fredholm determinants
\beq
{\rm Det}(1-z \hat{K}_{\star})\,,\quad \star=s,+
\eeq
where
\beq
\hat{K}_{\star} \cdot f(u) \equiv \int dv\, K_{\star}(u,v) f(v) \comma
\eeq
and $K_{\star}(u,v)$ is defined by 
\beq
K_{\star} (u,v) \equiv \frac{ \mathcal{K}_{\star}(u,v)}{ \sqrt{1+e^{\epsilon(u)}}\sqrt{1+e^{\epsilon(v)}}}\period
\eeq
For the kernel $\mathcal{K}_{s}$, the steps are completely analogous to the Tracy and Widom proof \cite{Tracy:1995ax}. The result for the kernel $\mathcal{K}_{+}$ is new as far as we know. 

\paragraph{Baxter-like functional equation}
For convenience, we will redefine here the auxiliary functions $\phi_{\star \, j}$ as
\beq
\psi_{\star\, j} (u)\equiv \frac{\sqrt{1+e^{\epsilon(u)}}}{\sqrt{2}} E_{\star}(u) \, \phi_{\star j} (u)\comma
\eeq
so that the recursion obeyed by $\psi_j$ now reads
\beq \label{eq:recpsi}
\psi_{\star \, j}(u) = \int dv\,\frac{\mathcal{K}_{\star} (u,v)}{1+e^{\epsilon(v)}} \, \psi_{\star \, j-1}(v)\comma \quad \text{with}\quad \psi_{\star \, 0} (u)\equiv \frac{\sqrt{1+e^{\epsilon(u)}}}{\sqrt{2}} E_{\star}(u) \period
\eeq
We now convert this representation into a functional equation by using the identity (\ref{eq:coshid}),
\beq
\psi_{\star \, j}^{++}+ \psi_{\star \, j}^{--} = \frac{2\pi \delta_{\star} }{1+e^{\epsilon}}\,  \psi_{\star\,  j-1}\comma
\eeq
where $\delta_{s}=1$ and $\delta_{+}=2$ and we have used that $\psi_{+ j}(-u)=\psi_{+ j}(u)$.  
A more useful recursion relation for what follows can be obtained by defining new functions $P_{\star}(u)$ and $Q_{\star}(u)$ out of the $\psi_{\star\, j}$ (we have also written them in (\ref{eq:PQmt}) in terms of $\phi_{\star \,j}$),
\beq \label{eq:PQ}
P_{\star}(u) \equiv \frac{\sqrt{2} }{\sqrt{1+e^{\epsilon(u)}}}\sum_{j=0}^{\infty} z^{2j+1} \psi_{\star\, 2j+1}(u)\comma \quad Q_{\star}(u) \equiv \frac{\sqrt{2} }{\sqrt{1+e^{\epsilon(u)}}}\sum_{j=0}^{\infty} z^{2j} \psi_{\star \, 2j}(u) \comma
\eeq
which immediately allows us to arrive at the Baxter-like equations (\ref{eq:baxterPQmt}), provided we use that $\epsilon(u)$ is an $i\pi$-periodic function.
We can easily determine the asymptotic behavior of these functions directly from their definition (\ref{eq:PQ}). In fact, we have that $\lim_{u\rightarrow \infty}\psi_j(u) =0$ for $j>0$ while for $j=0$ the asymptotics are  $\lim_{u\rightarrow \infty} \psi_{0}(u) =  \sqrt{\frac{1+e^{\epsilon(u)}}{2}} E_{\star}(u)$. With this we obtain that,
\beq \label{eq:asympt}
\lim_{u\rightarrow \infty} \frac{P_{\star}(u)}{E_{\star}(u)} = 0 \quad\quad \lim_{u \rightarrow \infty} Q_{\star}(u) = E_{\star}(u) \period
\eeq

\paragraph{Useful identity} We will now prove an identity that will play an important role in the derivation,
\beq \label{eq:proposition}
Q_{\star}^{+}Q_{\star}^{-}-P_{\star}^{+}P_{\star}^{-}= E_{\star}^{+}E_{\star}^{-}\period
\eeq
We  follow similar steps as in \cite{Tracy:1995ax}. Let us start by defining the functions $S_{\star}(u)$ as 
\beq
S_{\star} \equiv Q_{\star}^{+}Q_{\star}^{-} -P_{\star}^{+} P_{\star}^{-}\period
\eeq
From the equations (\ref{eq:baxterPQmt}) it follows that $S_{\star}^{-} = -S_{\star}^{+}$ and therefore $S_{\star}(u)$ is $i \pi/2$-periodic. From the definition of $S_{\star}(u)$ one can show that this function is analytic in the strip $[-i\pi/2, i\pi/2]$ (a  rigorous proof of all these statements can be found in \cite{Tracy:1995ax}). This then implies that the function
\beq \label{eq:zero}
1- \frac{S_{\star}}{E_{\star}^{+}E_{\star}^{-}}
\eeq
is analytic and bounded. For the case of $\star=+$, the apparent singularities at $u= i\pi/4 + \pi n$ with $n \in \mathbb{Z}$, are in fact cancelled by the numerator. Indeed, say at $u = i \pi/4$ we have that
\beq
S_{+}(i \pi/4) = Q_{+}(i\pi/2) Q_{+}(0) - P_{+}(i\pi/2) P_{+}(0) \period
\eeq
From the relation $Q_{+} = E_{+} +\hat{K}_{+} \hat{P}$ (which follows from the definitions) we have that
\beq
\hat{K}_{+} \cdot P_{+}(i\pi/2) = v_{+}(0) Q_{+}(0) \comma
\eeq
so that $Q_{+}(i\pi/2) =E_{+}(i\pi/2) +v_{+}(0) P_{+}(0)$. Analogously, from $P_{+} = \hat{K}_{+} \cdot Q_{+} $ we get that $P_{+}(i\pi/2)=v_{+}(0)  Q_{+}(0)$. Finally we then obtain that
\beq
\frac{S_{+}(i \pi/4)}{E_{+}(i \pi/4) E_{+}(0)} = \frac{Q_{+}(0)}{E_{+}(0)}\comma
\eeq
which is finite.
Given the asymptotics (\ref{eq:asympt}) we then conclude that the function (\ref{eq:zero}) is zero and the identity follows.

\paragraph{Final functional equations} We now have all the tools at hand to derive some functional equations involving  $R_{{\rm{e}}\,\star}$ and $R_{{\rm{o}}\,\star}$ which can be expressed in term of $P_{\star}$ and $Q_{\star}$ through (\ref{eq:RandPQ}). Again, once we determine these functions, we are able to compute each term in the series expansion of the Fredholm determinant (\ref{eq:detexpand}). The derivation of the functional equations works similarly for both kernels $\star=s,+$ with some small differences in explicit formulae and therefore we will derive them both simultaneously. 
As noted in \cite{Tracy:1995ax}, it proves useful to introduce an auxiliary function out of a particular combination of $P$ and $Q$, namely
\beq \label{eq:eta1}
\eta_{\star} -i \equiv - i \frac{(Q_{\star}^{+}-P_{\star}^{+})(Q_{\star}^{-}+P_{\star}^{-})}{E^{+}_{\star}E^{-}_{\star}}\period
\eeq
What we will show now is that with $\eta_{\star}$, $R_{{\rm{e}}\,\star}$ and $R_{{\rm{o}}\, \star}$ we can find a closed system of functional equations that we will later be able to invert and determine these three functions explicitly. 

Note that thanks to the identity (\ref{eq:proposition}) derived before, we can equally write (\ref{eq:eta1}) as
\beq \label{eq:eta2}
\eta_{\star}+i = i \frac{(Q_{\star}^{+}+P_{\star}^{+})(Q_{\star}^{-}-P_{\star}^{-})}{E^{+}_{\star}E^{-}_{\star}}\period
\eeq
The first functional equation is immediately obtained by taking the product of the previous two equations and taking the logarithm, getting
\beq \label{eq:fun1}
\log(1+\eta_{\star}^2) = \log(1+e^{\epsilon^{+}})+\log(1+e^{\epsilon^{-}})+\log \tilde{R}_{{\rm{e}}\,\star}^{+}+\log \tilde{R}_{{\rm{e}}\,\star}^{-}\comma
\eeq
where $\tilde{R}_{{\rm{e}}\,s}(u) =R_{{\rm{e}}\,s}(u) $ and $\tilde{R}_{{\rm{e}}\,+} (u)=\frac{\cosh(2u)}{\cosh(u)^2}R_{{\rm{e}}\,+}(u) $.
By considering the difference of the logarithmic derivatives of equations (\ref{eq:eta1}) and (\ref{eq:eta2}) we obtain yet another functional equation
\beq \label{eq:fun2}
\frac{2i\, \eta'_{\star}}{\eta_{\star}^2+1} = 2 i \arctan(\eta_{\star})' = \frac{\tilde{R}^{+}_{{\rm{o}}\,\star}}{\tilde{R}^{+}_{{\rm{e}}\,\star}}-\frac{\tilde{R}^{-}_{{\rm{o}}\,\star}}{\tilde{R}^{-}_{{\rm{e}}\,\star}} \comma
\eeq
where $\tilde{R}_{{\rm{o}}\,s}(u) =R_{{\rm{o}}\,s}(u) $ and $\tilde{R}_{{\rm{o}}\,+} (u)=\frac{\sinh(2u)}{\cosh(u)^2}R_{{\rm{e}}\,+}(u) $\,.
The last equation is again obtained from a combination of (\ref{eq:eta1}) and (\ref{eq:eta2}). By shifting the argument of (\ref{eq:eta1})  as $u\rightarrow u- \frac{i \pi}{4}$ and (\ref{eq:eta2}) as $u\rightarrow u +\frac{i \pi}{4}$ we obtain
\beq
\eta_{\star}^{-}-i= i \frac{ (Q_{\star}-P_{\star})(Q_{\star}^{--}+P_{\star}^{--})}{E_{\star} E_{\star}^{++}} \comma
\eeq 
and
\beq
\eta_{\star}^{+}-i= i \frac{ (Q_{\star}^{++}+P_{\star}^{++})(Q_{\star}-P_{\star})}{E_{\star} E_{\star}^{++}}\comma
\eeq
where we used $E^{++}= -E^{--}$ which itself follows from the periodicity $\epsilon^{++}=\epsilon^{--}$. We then sum these two equations and use the relation (\ref{eq:baxterPQmt}) to arrive at
\beq \label{eq:fun3}
\eta^{+}_{\star} +\eta^{-}_{\star} = 2 \pi \delta_{\star}\left( \frac{\cosh(u)}{\sinh(u)} \right)^{\delta_{\star}-1}\, \tilde{R}_{{\rm{e}}\,\star}
\eeq
where we remind that $\delta_{s}=1$ and $\delta_{+}=2$. 

\section{Details on the comparison with Liouville boundary data} \label{app:liouville}
In this appendix, we elaborate on the comparison with the Liouville bulk one-point function. We have expressed the $g$-function in terms of a momentum $P$ which is defined as a function of $L$ by the quantum version of (\ref{eq:quantization}) (see \cite{Zamolodchikov:1995aa}),
\beq
\frac{1}{i} \log \left(-S(P)\right)-4 P \left(b+\frac{1}{b}\right) \log\left( \frac{L}{2\pi}\right)=\pi 
\eeq
where $S(P)$ is the reflection matrix
\beq
S(P)= - \left( \pi  \mu \gamma(b^2) \right)^{-2i P/b} \frac{\Gamma(1+2 i P/b) \Gamma(1+2 i P b)}{\Gamma(1-2 i P/b) \Gamma(1-2 i P b)} \period
\eeq
The cosmological constant $\mu$ is related to the sinh-Gordon mass by
\beq \label{eq:mass}
\pi \mu\gamma(b^2)  = \left( \frac{m}{8 \sqrt{\pi}} p^{p} (1-p)^{1-p} \Gamma\left(\frac{p}{2}\right)\Gamma\left(\frac{1-p}{2}\right)\right)^{2 +2 b^2}\;\;\; {\rm{with}}\;\;\; p=\frac{b^2}{1+b^2}\,.
\eeq
For the self-dual sinh-Gordon studied in this paper, we express the results in terms of the mass using (\ref{eq:mass}) and finally set $b=1$. Another practical way of finding the relation between momentum and the sinh-Gordon radius $L$ has been applied in \cite{Zamolodchikov:1995aa} and uses the bulk TBA,
\beq
\epsilon(u) = - m L \cosh(u)+ \mathcal{K}_s \ast \log(1+e^{-\epsilon})\comma
\eeq
from which we compute the effective central charge
\beq
c_{{\rm{eff}}} = \frac{3m L}{\pi^2} \int du \cosh(u) \log(1+e^{-\epsilon(u)})\period
\eeq
For small $L$, $c_{{\rm{eff}}}$ is approximately related to the momentum by
\beq
c_{{\rm{eff}}} = 1-24 P^2 +\mathcal{O}(L^2)\,,
\eeq
which we can use to go from $L$ to $P$ or vice-versa.
\section{Factorization of SoV integral in spin chains \label{ap:factorization}}
In this appendix, we apply the argument presented in section \ref{sec:sov} to the SU(2) (XXX) spin chain and reproduce an integral representation for the overlap between the boundary state and the Bethe state, which was first derived in appendix R of \cite{Jiang:2019xdz}. We also apply the same argument to the SL(2) spin chain and conjecture the SoV integral representation for the exact overlap \cite{Jiang:2019xdz,Jiang:2020sdw}.

\paragraph{SU(2) spin chain}
To simplify the discussion, we assume that the length of the spin chain $L$ is even and is twice the number of magnons $m$. The integral representation for more general cases can be derived from it by following the argument in appendix R of \cite{Jiang:2019xdz}.

 The starting point is the SoV integral for the norm \cite{Kazama:2013rya}
\beq\label{eq:apintSoV}
 \langle {\bf u}|{\bf u}\rangle \propto\oint_{\mathcal{C}}\left(\prod_{k=1}^{L}\frac{dx_k}{2\pi i}\frac{Q(x_k)Q(x_k)}{Q_{\theta}^{+}(x_k)Q_{\theta}^{-}(x_k)}\right)\prod_{1\leq j<k\leq L}(x_j-x_k)\left(e^{2\pi x_j}-e^{2\pi x_k}\right)\comma
\eeq
where $|{\bf u}\rangle$ is the onshell Bethe state with rapidities ${\bf u}=\{u_1,\ldots,u_m\}$, while $Q(x)$ and $Q_{\theta}(x)$ are given by
\beq
Q(x)\equiv \prod_{j=1}^{m(=L/2)}(x-u_j)\comma\qquad Q^{\pm}_{\theta}(x)\equiv \prod_{k=1}^{L}(x-\theta_k\pm i/2)\comma
\eeq
with $\theta_j$ being the inhomogeneities. The contour $\mathcal{C}$ encircles all the poles in the integrand $\theta_j\pm \frac{i}{2}$.  

Now, up to a $\theta$-dependent prefactor, the integral \eqref{eq:apintSoV} coincides with the following slightly different integral
\beq\label{eq:apintSoV2}
 \langle {\bf u}|{\bf u}\rangle \propto\oint_{\mathcal{C}}\left(\prod_{k=1}^{L}\frac{dx_k}{2\pi i}\frac{Q(x_k)Q(x_k)e^{-\pi x_k}}{Q_{\theta}^{+}(x_k)Q_{\theta}^{-}(x_k)}\right)\prod_{1\leq j<k\leq L}(x_j-x_k)\sinh (\pi (x_j-x_k))\period
\eeq
The equivalence between the two follows from the identity
\beq
\prod_{k}e^{-\pi x_k}\prod_{j<k}\sinh (\pi (x_j-x_k))=\underbrace{\left(\prod_{k}e^{-2\pi \frac{L}{2}x_k}\right)}_{\equiv\,\, {\tt f}}\,\,\prod_{j<k}\left(e^{2\pi x_j}-e^{2\pi x_k}\right)\comma
\eeq
and the fact that the extra factor ${\tt f}$ gives a $\theta$-dependent overall constant $\prod_k (-1)^{\frac{L}{2}}\exp \left(\pi L \theta_k\right)$, regardless of which poles in the integrand we choose to pick\footnote{Owing to the Vandermonde factor $(x_j-x_k)(e^{2\pi x_j}-e^{2\pi x_k})$, we can only pick one from each set $\{\theta_k+\frac{i}{2}, \theta_k-\frac{i}{2}\}$. This leaves us $2^{L}$ choices of poles (up to permutations). It is straightforward to verify that ${\tt f}$ gives the same answer $\prod_k (-1)^{\frac{L}{2}}\exp \left(\pi L \theta_k\right)$ for all such choices.}.

Using the Vandermonde determinant formula, we can rewrite \eqref{eq:apintSoV2} into a determinant of a $L\times L$ matrix
\beq
\langle {\bf u}|{\bf u}\rangle \propto \det \left(M_{j,k}\right)_{1\leq j,k\leq L}\comma
\eeq
with
\beq\label{eq:apMdef}
M_{j,k}\equiv \oint_{\mathcal{C}}\frac{dx}{2\pi i}\frac{Q(x)Q(x)}{Q_{\theta}^{+}(x)Q_{\theta}^{-}(x)}x^{j-1}e^{\pi (2k-L-2)x}\period 
\eeq

To proceed, we assume that the state is parity-symmetric, namely $Q(x)=Q(-x)$ and $Q_{\theta}^{+}(x)=Q_{\theta}^{-}(-x)$. We can then rewrite \eqref{eq:apMdef} as
\beq\label{eq:SU2coshsinh}
M_{j,k}\equiv \oint_{\mathcal{C}}\frac{dx}{2\pi i}\frac{Q(x)Q(x)}{Q_{\theta}^{+}(x)Q_{\theta}^{-}(x)}x^{j-1}\times \begin{cases}\cosh\left(\pi (2k-L-2) x\right)\quad &j:\text{even}\\\sinh\left(\pi (2k-L-2) x\right)\quad &j:\text{odd} \end{cases}\period
\eeq
In particular $M_{j,\frac{L+2}{2}}=0$ for odd $j$.
Given this structure, it is natural to re-group $j$ into even and odd integers and $k$ into $\{\frac{L}{2}+1,\ldots, L\}$ and $\{\frac{L}{2},\ldots, 1\}$. As a result, we get the following block-matrix structure
\beq
M=\pmatrix{cc|cc}{a&A&A&a^{\prime}\\\hline 0&B&-B&-b^{\prime}}\comma
\eeq
where $A$ and $B$ are $\frac{L}{2}\times (\frac{L}{2}-1)$ matrices, and $a$, $a^{\prime}$ and $b^{\prime}$ are $\frac{L}{2}$-component vectors:
\beq
\begin{aligned}
&A_{j,k}\equiv M_{2j,k+\frac{L+2}{2}}=M_{2j,-k+\frac{L+2}{2}}\comma\qquad B_{j,k}\equiv M_{2j-1,k+\frac{L+2}{2}}=-M_{2j-1,-k+\frac{L+2}{2}}\comma\\
&a_j\equiv M_{2j,\frac{L+2}{2}}\comma\quad a^{\prime}_j\equiv M_{2j,1}\comma\quad b^{\prime}\equiv -M_{2j-1,1}\comma\qquad (1\leq j\leq \tfrac{L}{2}\comma\quad 1\leq k\leq \tfrac{L}{2}-1)\period
\end{aligned}
\eeq
We can then show that the determinant factorizes as
\beq
\begin{aligned}
\det M &=\det \pmatrix{cc|cc}{a&A&A&a^{\prime}\\\hline 0&B&-B&-b^{\prime}} =\det \pmatrix{cc|cc}{a&2A&A&a^{\prime}\\\hline 0&0&-B&-b^{\prime}}\\
&=2^{\frac{L}{2}-1}(-1)^{\frac{L}{2}}\det \tilde{A} \det \tilde{B}\comma
\end{aligned}
\eeq
where $\tilde{A}$ and $\tilde{B}$ are square matrices of size $L/2$, 
\beq
\begin{aligned}
\tilde{A}_{j,k}&\equiv \oint_{\mathcal{C}}\frac{dx}{2\pi i}\frac{Q(x)Q(-x)}{Q_{\theta}^{+}(x)Q_{\theta}^{-}(x)}x^{2j-1}\cosh \left(2\pi (k-1)x\right)\comma\\
\tilde{B}_{j,k}&\equiv \oint_{\mathcal{C}}\frac{dx}{2\pi i}\frac{Q(x)Q(-x)}{Q_{\theta}^{+}(x)Q_{\theta}^{-}(x)}x^{2(j-1)}\sinh \left(2\pi kx\right)\period
\end{aligned}
\eeq
Here we replaced $Q(x)Q(x)$ with $Q(x)Q(-x)$ using the parity symmetry. 

From the analogy with the Gaudin norm, it is reasonable to conjecture that one of the determinants, $\det \tilde{A}$ or $\det \tilde{B}$, describes the overlap between the boundary state and the Bethe state. Applying the argument in section \ref{subsec:conjectureg} and using the Baxter equation
\beq
T(x)Q(x)=Q^{-}_{\theta}(x)Q(x+i)+Q^{+}_{\theta}(x)Q(x-i)\comma
\eeq 
one can show that $\det \tilde{A}$ reproduces the selection rule; namely it vanishes unless the $Q$-function is parity-symmetric. Therefore, we expect $\det \tilde{A}$ is the one that corresponds to the overlap. 

To make contact with the result in \cite{Jiang:2019xdz}, we need to rewrite $\det \tilde{A}$ into multiple integrals. To do so, we use the fact that $2\cosh (\alpha n x)$ is a polynomial of $2\cosh \alpha x$ of degree $n$,
\beq
2\cosh (\alpha n x)=(2\cosh \alpha x)^{n}+\cdots\period
\eeq
We can then add and subtract rows and columns, and replace $\tilde{A}$ with the following matrix without changing its determinant
\beq
\tilde{A}_{j,k}\mapsto \oint_{\mathcal{C}}\frac{dx}{4\pi i}\frac{Q(x)Q(-x)}{Q_{\theta}^{+}(x)Q_{\theta}^{-}(x)}x^{2j-1}\left(2\cosh (2\pi x)\right)^{k-1}\period
\eeq
We can then use the Vandermonde formula to rewrite it as the following multiple integral,
\beq
\det \tilde{A}\propto \oint_{\mathcal{C}}\left(\prod_{k=1}^{L}\frac{x_kdx_k}{2\pi i}\frac{Q(x_k)Q(-x_k)}{Q_{\theta}^{+}(x_k)Q_{\theta}^{-}(x_k)}\right)\prod_{j<k}\bar{\Delta} (x_j,x_k)\comma
\eeq
with 
\beq\label{eq:bardelta}
\bar{\Delta}(x,y)=(x^2-y^2)\left(2\cosh (2\pi x)-2\cosh (2\pi y)\right)=(x^2-y^2)\left(\sinh^2 (\pi x)-\sinh^2 (\pi y)\right)\period
\eeq
This is in agreement\footnote{Up to an overall factor which we did not keep track of.} with (R.21) in \cite{Jiang:2019xdz}, confirming our expectation.
\paragraph{Conjecture for SL(2) spin chain} The same argument applies also to the non-compact SL(2) spin chain. The SoV-integral representation for the norm was written down in \cite{Derkachov:2002tf} (see also (G.43) of \cite{Jiang:2016ulr}), and it reads
\beq\label{eq:SL2SoV}
\langle {\bf u}|{\bf u}\rangle\propto \int^{\infty}_{-\infty}\left(\prod_{k=1}^{L-1}\frac{dx_k}{2\pi}\frac{Q(x_k)Q(x_k)}{(\cosh \pi x_k)^{L}}\right)\prod_{j<k}\sinh (\pi (x_j-x_k))(x_j-x_k)\comma
\eeq
where $L$ is the length of the spin chain which we assume to be even while $Q(x)$ is the $Q$-function
\beq
Q(x)\equiv \prod_{s=1}^{m}(x-u_s)\comma
\eeq 
with $u_s$ being the rapidities of magnons. As with the SU(2) spin chain, the integral \eqref{eq:SL2SoV} can be converted into a determinant using the Vandermonde formula,
\beq
\langle {\bf u}|{\bf u}\rangle\propto \det (M_{j,k})_{1\leq j,k\leq L-1}\comma
\eeq
with
\beq
M_{j,k}\equiv \int_{-\infty}^{\infty}\frac{dx}{2\pi}\frac{Q(x)Q(x)}{(\cosh \pi x)^{L}}x^{j-1}e^{2\pi (k-\frac{L}{2})}\period
\eeq
If the state is parity-symmetric, one can rewrite this as\footnote{Note that the roles of even and odd $j$'s are swapped here as compared to \eqref{eq:SU2coshsinh} since the contour $\mathcal{C}$ in \eqref{eq:SU2coshsinh} is invariant under the parity $x\to -x$ while the contour here changes the orientation.}
\beq
M_{j,k}\equiv \int_{-\infty}^{\infty}\frac{dx}{2\pi}\frac{Q(x)Q(-x)}{(\cosh \pi x)^{L}}x^{j-1}\times \begin{cases}\cosh\left(2\pi (k-\tfrac{L}{2})x\right)\qquad &j:\text{odd}\\\sinh\left(2\pi (k-\tfrac{L}{2})x\right)\qquad &j:\text{even}\end{cases}\period
\eeq
We can then use the same trick as in the SU(2) spin chain; namely we reorder, add and subtract rows and columns. As a result, we obtain the following factorized expression:
\beq
\det M\propto \det \tilde{A}\times \det \tilde{B}\comma
\eeq
where $\tilde{A}$ and $\tilde{B}$ are the matrices of size $\tfrac{L}{2}-1$ and $\tfrac{L}{2}$ respectively and read
\beq
\begin{aligned}
&\tilde{A}_{j,k}\equiv \int_{-\infty}^{\infty}\frac{dx}{2\pi}\frac{Q(x)Q(-x)}{(\cosh \pi x)^{L}}x^{2j-1}\sinh\left(2\pi k x\right)\comma\\
&\tilde{B}_{j,k}\equiv \int_{-\infty}^{\infty}\frac{dx}{2\pi}\frac{Q(x)Q(-x)}{(\cosh \pi x)^{L}}x^{2(j-1)}\cosh\left(2\pi (k-1)x\right)\period
\end{aligned}
\eeq
Among the two determinants $\det \tilde{A}$ and $\det \tilde{B}$, the one that reproduces the selection rule is $\det \tilde{A}$. We thus expect that the overlap is given by $\det\tilde{A}$. To convert it into multiple integrals, we use the identity \eqref{eq:coshsinhidentity} and the Vandermonde formula. As a result, we get
\beq\label{eq:SL2final}
\begin{aligned}
&\det \tilde{A}\propto \int^{\infty}_{-\infty}\left(\prod_{k=1}^{\frac{L}{2}-1}\frac{x_k \sinh (2\pi x_k) dx_k}{2\pi}\frac{Q(x_k)Q(-x_k)}{(\cosh \pi x_k)^{L}}\right)\prod_{1\leq j<k\leq \frac{L}{2}-1}\bar{\Delta}(x_j,x_k)\period
\end{aligned}
\eeq
with $\bar{\Delta}(x,y)$ given by \eqref{eq:bardelta}.
We conjecture that \eqref{eq:SL2final} gives the SoV representation for the overlap between the boundary state and the Bethe state. Of course, to use this formula in practice, we need to determine a constant of proportionality. In principle, this should be doable by comparing \eqref{eq:SL2final} and the results in \cite{Jiang:2019xdz,Jiang:2020sdw}. It would be an interesting future problem to carry this out and prove our conjecture \eqref{eq:SL2final} (and correct it if needed).
\bibliographystyle{JHEP}
\bibliography{DraftRef}
\end{document}